\documentclass[aps, prd, twocolumn,showkeys, showpacs,amsmath,amssymb,superscriptaddress]{revtex4-1}
\usepackage{graphicx}
\usepackage{dcolumn}
\usepackage{bm}
\usepackage{epsfig}
\usepackage{subfigure}
\usepackage{color}
\usepackage{epstopdf}

\begin{document}
\title{Lattice Boltzmann model for ultra-relativistic flows}

\author{F. Mohseni} \email{mohsenif@ethz.ch} \affiliation{ ETH
  Z\"urich, Computational Physics for Engineering Materials, Institute
  for Building Materials, Schafmattstrasse 6, HIF, CH-8093 Z\"urich
  (Switzerland)}

\author{M. Mendoza} \email{mmendoza@ethz.ch} \affiliation{ ETH
  Z\"urich, Computational Physics for Engineering Materials, Institute
  for Building Materials, Schafmattstrasse 6, HIF, CH-8093 Z\"urich
  (Switzerland)}

\author{S. Succi} \email{succi@iac.cnr.it} \affiliation{Istituto per
  le Applicazioni del Calcolo C.N.R., Via dei Taurini, 19 00185, Rome
  (Italy),\\and Freiburg Institute for Advanced Studies,
  Albertstrasse, 19, D-79104, Freiburg, (Germany)}

\author{H. J. Herrmann}\email{hjherrmann@ethz.ch} \affiliation{ ETH
  Z\"urich, Computational Physics for Engineering Materials, Institute
  for Building Materials, Schafmattstrasse 6, HIF, CH-8093 Z\"urich
  (Switzerland)} \affiliation{Departamento de F\'isica, Universidade
  Federal do Cear\'a, Campus do Pici, 60455-760 Fortaleza, Cear\'a,
  (Brazil)}

\date{\today}
\begin{abstract}
  We develop a relativistic lattice Boltzmann model capable of
  describing relativistic fluid dynamics at ultra-high velocities, with
  Lorentz factors up to $\gamma \sim 10$. 
  To this purpose, we first build a new lattice kinetic scheme by
  expanding the Maxwell-J\"uttner distribution function 
  in an orthogonal basis of polynomials and applying an
  appropriate quadrature, providing the discrete versions of the
  relativistic Boltzmann equation and the equilibrium distribution. 
  To achieve ultra-high velocities, we include a flux 
  limiter scheme, and introduce the bulk viscosity by a suitable
  extension of the discrete relativistic Boltzmann equation. 
  The model is validated by performing simulations of
  shock waves in viscous quark-gluon plasmas and comparing with 
  existing models, finding very good agreement. 
  To the best of our knowledge, we for the first time successfully simulate viscous
  shock waves in the highly relativistic regime. Moreover, we show that
  our model can also be used for near-inviscid flows even at very high
  velocities. Finally, as an astrophysical application, we simulate a relativistic
  shock wave, generated by, say, a supernova explosion, colliding with a massive 
  interstellar cloud, e.g. molecular gas. 
\end{abstract}


\maketitle
\section {Introduction}

Relativistic fluid dynamics plays an important role in many contexts
of astrophysics and high-energy physics, e.g. jets emerging from the
core of galactic nuclei or gamma-ray bursts \cite{relativisticjet},
shock induced Ritchmyer-Meshkov instabilities
\cite{nishihara2010richtmyer} and quark-gluon plasmas produced in
heavy-ion collisions \cite{shuryak2004does}.  Hence, various numerical
methods have been developed to study the relativistic hydrodynamics.
Most of these methods are focussed on the solution of the corresponding
relativistic macroscopic conservation equations. 
Among others, one can mention the methods based on second-order
Lax-Wendroff scheme\cite{dubal1991numerical}, smoothed particle
hydrodynamics techniques \cite{TREESPH,siegler2008smoothed}, Glimm’s
(random choice) method \cite{wen1997shock} and high resolution
shock-capturing methods \cite{Generalrelativistichydrodynamics}. 
Other methods, instead of solving the macroscopic equations, tackle the
problem from the microscopic and mesoscopic points of view
\cite{yang1997kinetic}.
To this regard, the lattice Boltzmann (LB) method
\cite{Benzi1992145,chen1992recovery,succi2001lattice} a relatively
new numerical approach, based on a minimal lattice version of the
Boltzmann kinetic equation, has enjoyed increasing popularity 
for the last two decades.
Within LB, representative particles stream and collide on the nodes 
of a regular lattice, with sufficient symmetry to
reproduce the correct equations of macroscopic hydrodynamics. 
The main highlights of LB are its computational simplicity, easy handling
of complex geometries, and high amenability to parallel computing
\cite{doi:10.1142/S0129183197000862}. 
The LB method has met with remarkable success for  
the simulation of a broad variety of complex
flows, from fully developed turbulence, all the way down to
nanoscale flows of biological interest
\cite{PhysRevLett.61.2332,higuera2007boltzmann,succi2008lattice} .


From a mathematical viewpoint,  the standard lattice
Boltzmann model can be obtained by expanding the equilibrium
distribution, i.e. Maxwell-Boltzmann distribution, in a Hermite
polynomials and using the nodes of polynomials, up to a certain order
as the corresponding discretized velocities
\cite{shan1998discretization}, using the Bhatnagar-Gross-Krook (BGK)
approximation for the collision operator \cite{PhysRev.94.511}.
While the applications of the LB scheme cover an impressive array
of complex fluid flows, its relativistic extension has been developed
only in last few years \cite{PhysRevLett.105.014502,rlbPRD}. 

The relativistic LB (RLB) model was constructed by expanding the
distribution function in powers of the fluid speed and finding the
corresponding coefficients (Lagrange multipliers), by matching the
moments of the Maxwell-J\"uttner distribution in continuum velocity
space.  This model was shown capable of simulating weakly and
moderately relativistic viscous flows, with $\beta =u/c \sim 0.3$, $u$
being the typical flow speed.  In particular, RLB was applied to the
simulation of shock waves in quark-gluon plasmas, showing very good
agreement with the results obtained by solving the full Boltzmann
equation for multi-parton scattering (BAMPS),
\cite{bouras2009relativistic},

However, the aforementioned matching procedure does not provide a
unique solution for the discrete equilibrium distribution function,
satisfying the hydrodynamics moments of the Maxwell-J\"uttner
distribution.  Moreover, the model lacks dissipation for the zero
component of the energy-momentum tensor, and imposes a non-physical
diffusion in the conservation of the number of particles \cite{mrtrlbPRD}.
These flaws, albeit very minor at moderate flow speeds, may become a
concern for strongly relativistic flows. 
It is therefore highly desirable to
develop more general and systematic approaches.
To this purpose, let us observe that, due to the non-separability of the
Maxwell-J\"uttner distribution function into the three components of
the momentum in Cartesian coordinates, its expansion in orthogonal
polynomials is not as natural as in the classical case and some
deliberation is required. For the fully relativistic regime,
neglecting particle masses, and by using spherical coordinates, a
lattice Boltzmann algorithm for the relativistic Boltzmann equation
was developed in Ref. \cite{PhysRevC.84.034903}. In this paper, the
Maxwell-J\"uttner distribution function was expanded in an orthogonal
polynomials basis and discretized using a Gauss quadrature
procedure. The model was based on the Anderson-Witting collision
operator \cite{Anderson1974466}.  The results of simulating viscous
quark-gluon plasma were compared to other hydrodynamic simulations and
very good agreement was observed. However, using spherical coordinates
makes the scheme incompatible with a cartesian lattice, and consequently, in the
streaming procedure, a linear interpolation is required at each time
step. Therefore, some crucial properties of the classical LB,
e.g. exact streaming (zero numerical dispersion) and negative
numerical diffusivity, are lost in the process.

In this paper, we develop a relativistic lattice Boltzmann model by
expanding the Maxwell-J\"uttner distribution in a set of orthogonal
polynomials, and performing an appropriate quadrature in order to
adjust the scheme to a D3Q19 (19 discrete velocities in three spatial
dimensions) cell configuration \cite{PhysRevLett.105.014502, rlbPRD}.
Moreover, we extend the model by using a minimum modulus flux limiter
scheme and introducing the bulk viscosity term into the Boltzmann
equation. We show that the model is numerically stable also at very
high velocities, i.e. Lorentz factors up to $\gamma \sim 10$.
Additionally, we show that this model can also be used to simulate
near-inviscid flows, which corresponds to solve the Euler equation on
the macroscopic level. This is well suited for astrophysical
applications, where the viscosity is usually negligible.  
In fact, the astrophysical context presents possibly the richest arena 
for future applications of the present RLB scheme.

The paper is organized as follows: in Sec.~\ref{Model Description},
the model description is presented in detail; and in
Sec.~\ref{Validation and Results}, several validations with other
existing numerical models along with some results for shock waves in
viscous quark-gluon plasmas and a 3D simulation of a shock wave
colliding with a massive interstellar cloud are presented. Finally, in
Sec.~\ref{Conclusions}, a discussion about the model and the results
is provided.

\section {Model Description}
\label {Model Description}

We start the description of our model by writing the Maxwell-J\"uttner
equilibrium distribution function as
\begin{equation}\label{eqmj}
  f^{\rm eq} = A \exp(-p_\mu U^\mu/k_B T),
\end{equation}
where, $A$ is a normalization constant, $k_B$ the Boltzmann constant,
$T$ the temperature and $(p^\mu) = (E/c, \boldsymbol{p})$ is the
4-momentum, with the energy of the particles $E$ defined by
\begin{equation}
  E=cp^0=\frac{mc^2}{\sqrt{1-u^2/c^2}},
\end{equation}
The macroscopic 4-velocity is $(U^\mu) = (c, \boldsymbol{u})
\gamma (u)$, with $\boldsymbol{u}$ the three-dimensional velocity,
$\gamma(u)=1/\sqrt{1-u^2/c^2}$ the Lorentz's factor, $m$ the mass, and
$c$ the speed of light. The relativistic Boltzmann equation, based on
the Marle collision operator \cite{marle}, reads as follows
\begin{equation}\label{eqboltzmann}
  p^\mu \partial_\mu f = -\frac{m}{ \tau}( f - f^{\rm eq} ),
\end{equation}
where $f$ is the probability distribution function, and $\tau$ the
single relaxation time.
It is possible to write the Maxwell-J\"uttner distribution in a
simpler form by introducing the following change of variables:
\begin{equation}
  \xi^\mu = \frac{p^\mu/m}{c_s} , \quad \chi^\mu = \frac{U^\mu}{c_s},
\end{equation}
\begin{equation}
  c_s = \sqrt{\frac{k_B T}{m}} , \quad \nu = c/c_s \quad ,
\end{equation}
and therefore, by replacing the new variables in Eq. \eqref{eqmj} we
have
\begin{equation}
f^{\rm eq} = A \exp(-\xi_\mu \chi^\mu) .
\end{equation}
The temporal components, $\xi^0 $ and $\chi^0$, can be calculated by
the relations
\begin{equation}
\xi^0 = \sqrt{|\boldsymbol {\xi}|^2+\nu^2},
\end{equation}
\begin{equation}
\chi^0 = \nu \gamma(u), \quad \gamma(u)=\sqrt{1+\frac{|\boldsymbol{\chi}|^2}{\nu^2}}.
\end{equation}

In analogy to the classical procedure of expanding the Maxwell-Boltzmann
distribution in Hermite polynomials, we can also expand the
Maxwell-J\"uttner distribution, using orthogonal polynomials of the
following form:
\begin{equation}\label{eqexpansion}
f ^{\rm eq}(\boldsymbol{\xi}, \boldsymbol{x}, t) = w(\boldsymbol{\xi}) \sum_{n=0}^{\infty} \frac{a_{(n)}(\boldsymbol{x}, t)}{N_{(n)}}F_{(n)}(\boldsymbol{\xi}) ,
\end{equation}
where
\begin{equation}
\int w(\boldsymbol{\xi}) F_{(n)}(\boldsymbol{\xi}) F_{(m)}(\boldsymbol{\xi}) \,\frac{d^3\xi}{\xi^0} = 0,
\end{equation}
for $m\not=n$, and
\begin{equation}
N_{(n)} = \int  w F_{(n)} F_{(n)} \,\frac{d^3\xi}{\xi^0}.
\end{equation}
To construct the appropriate orthogonal polynomials, we
introduce the corresponding weight function as the equilibrium
distribution at the local rest frame,
\begin{equation}
w(\boldsymbol{\xi}) = A \exp(-\nu \xi^0) .
\end{equation}

Using the procedure proposed by Stewart \cite{stewart}, where the
non-equilibrium distribution was expanded around the equilibrium, and
the Maxwell-J\"uttner distribution was used as the weight function to
find the orthogonal polynomials, we can take up to second order,
\begin{equation}
  F_{(0)} = 1 ,
\end{equation}
and 
\begin{equation}
  F_{(1)}^\alpha = \xi^\alpha - a^\alpha,
\end{equation}
\begin{equation}
  F_{(2)}^{\alpha \beta}=\xi^\alpha \xi^\beta - a^{\alpha \beta}_\gamma  F_{(1)}^\gamma - b^{\alpha \beta}, 
\end{equation}
where $a^\alpha$, $a^{\alpha \beta}$ and $b^{\alpha \beta}$ are
unknowns to be calculated using the Gram-Schmidt orthogonalization
procedure
\begin{eqnarray}
\int w  F_{(0)} F_{(1)}^\alpha \,\frac{d^3\xi}{\xi^0} &=& \int w  F_{(0)} F_{(2)}^{\alpha \beta} \,\frac{d^3\xi}{\xi^0} \nonumber \\
&=& \int w  F_{(1)}^\alpha F_{(2)}^{\alpha\beta} \,\frac{d^3\xi}{\xi^0} = 0 .
\end{eqnarray}
The normalization coefficient for each polynomial is given by
$\sqrt{N_{(n)}}$, and the coefficient $a_{(n)}$ is calculated using the
relation
\begin{equation}
a_{(n)} = \int f^{\rm eq} F_{(n)} \,\frac{d^3\xi}{\xi^0} .
\end{equation}

To calculate the coefficients $a^\alpha$, $a^{\alpha \beta}$ and
$b^{\alpha \beta}$, one needs the moments of the Maxwell-J\"uttner
distribution, which up to second order are given by \cite
{cercignani}
\begin{equation}\label{moment1}
\int f^{\rm eq} \,\frac{d^3\xi}{\xi^0} = 4\pi A K_1(\nu^2),
\end{equation}
\begin{equation}\label{moment2}
\int \xi^\alpha f^{\rm eq} \,\frac{d^3\xi}{\xi^0} = 4\pi A K_2(\nu^2)\chi^\alpha,
\end{equation}
\begin{equation}\label{moment3}
\int \xi^\alpha \xi^\beta f^{\rm eq} \,\frac{d^3\xi}{\xi^0} = -4\pi A \left(K_2(\nu^2)\eta^{\alpha\beta} - K_3(\nu^2)\chi^\alpha\chi^\beta\right),
\end{equation}
where $K_n(\nu^2)$ is the modified Bessel function of the second kind
of order $n$ and $\eta^{\alpha\beta}$ is the Minkowski metric tensor
with the signature $(+,-,-,-)$. The moments with respect to the weight
function can be determined by considering the above integrals in the
local Lorentz rest frame.

For the sake of simplicity we define $\phi^\alpha$ as
\begin{equation}
(\phi^\alpha)=(\chi^0, \boldsymbol{0}),
\end{equation}
and by using the orthogonalization relations to calculate the
unknowns, the resulting relativistic orthogonal polynomials are given
by
\begin{equation}
F_{(0)}=1,
\end{equation}
\begin{equation}
F_{(1)}^\alpha = \xi^\alpha -\frac{K_2(\nu^2)}{K_1(\nu^2)} \phi^\alpha ,
\end{equation}
\begin{equation}
F_{(2)}^{\alpha \beta} = \xi^\alpha \xi^\beta - a^{\alpha \beta}_\gamma  F_{(1)}^\gamma - b^{\alpha \beta} ,
\end{equation}
where
\begin{equation}
b^{\alpha \beta} = \frac{K_3(\nu^2)}{K_1(\nu^2)}\phi^{\alpha \beta} -  \frac{K_2(\nu^2)}{K_1(\nu^2)}\eta^{\alpha \beta},
\end{equation}
and
\begin{equation}
\begin{array}{c}
a^{\alpha \beta}_\gamma = \frac{\eta_{\gamma\delta}+D(\nu) \phi_\gamma\phi_\delta} {2 K_2(\nu^2) D(\nu)} \bigg[K_3(\nu^2)\left( \eta^{\alpha\delta} \phi^\beta+\eta^{\beta\delta} \phi^\alpha \right)  \\ 
-\left(K_3(\nu^2) - \frac{[K_2(\nu^2)]^2}{K_1(\nu^2)} \right) \eta^{\alpha \beta}\phi^\delta \\ 
 +\left(K_4(\nu^2)-\frac{K_2(\nu^2)
     K_3(\nu^2)}{K_1(\nu^2)}\right)\phi^\alpha\phi^\beta\phi^\delta\bigg]
 .
\end{array}
\end{equation}
Here, the function $ D(\nu)$ is defined by
\begin{equation}
[1+D(\nu)]^{-1}= 1+ \frac{K_2(\nu^2)}{K_1(\nu^2)}-\frac{K_3(\nu^2)}{K_2(\nu^2)} .
\end{equation}

By following this procedure, we can calculate higher
  order polynomials. However, since in this work we
  are interested in recovering only up to the second moment of the
  Maxwell-J\"uttner distribution (energy-momentum tensor), using the
  expansion up to the second order is sufficient. In particular, the
  third, fourth and fifth order moments, which are needed to describe
  highly viscous fluids, would increase dramatically the complexity of
  our expansion, and consequently its numerical implementation. 
  This is a very interesting subject for future extensions of this works.

Additionally, in the ultrarelativistic limit, where $k_BT\gg mc^2 $,
i.e. $\nu \ll 1$, we can use the following asymptotic relation:
\begin{equation}\label{eqlimit}
\lim_{\nu\to 0} K_n(\nu^2) = \frac{2^{n-1}(n-1)!}{\nu^{2n}} .
\end{equation}

Using the resulting polynomials $F_{(n)}$, coefficients $a_{(n)}$ and
$N_{(n)}$ with Eqs.~\eqref{eqexpansion} and \eqref{eqlimit} we can
expand the Maxwell-J\"uttner distribution in orthogonal polynomials,
\begin{equation}
\begin{array}{c}
  f^{\rm eq} \simeq Ae^{-\nu\xi^0}\bigg\{   1+ \left( \frac{\chi^0\xi^0+3}{2}-\frac{\chi^0}{\nu}-\frac{\xi^0\nu}{4}\right)(\boldsymbol{\xi}.\boldsymbol{\chi}) \\
  + \xi^x\xi^y\chi^x\chi^y+\xi^x\xi^z\chi^x\chi^z+\xi^y\xi^z\chi^y\chi^z \\ 
  +\frac{4}{\nu^4-6\nu^2-15} 
  \big[(\xi^{x})^2(\chi^{x})^2+(\xi^{y})^2(\chi^{y})^2+(\xi^{z})^2(\chi^{z})^2 \\
  +\left(\frac{1-\nu^2}{\nu}\xi^0-\frac{4-2\nu^2}{\nu^2}\right)(\boldsymbol{\chi}.\boldsymbol{\chi)}\big]  \bigg\},
\end{array}
\end{equation}
up to second order and in the ultrarelativistic regime.

\begin{figure}
\begin{center}
\includegraphics [trim=0mm 0mm 0mm 0mm, clip, width=0.9\columnwidth, height=0.7\columnwidth]{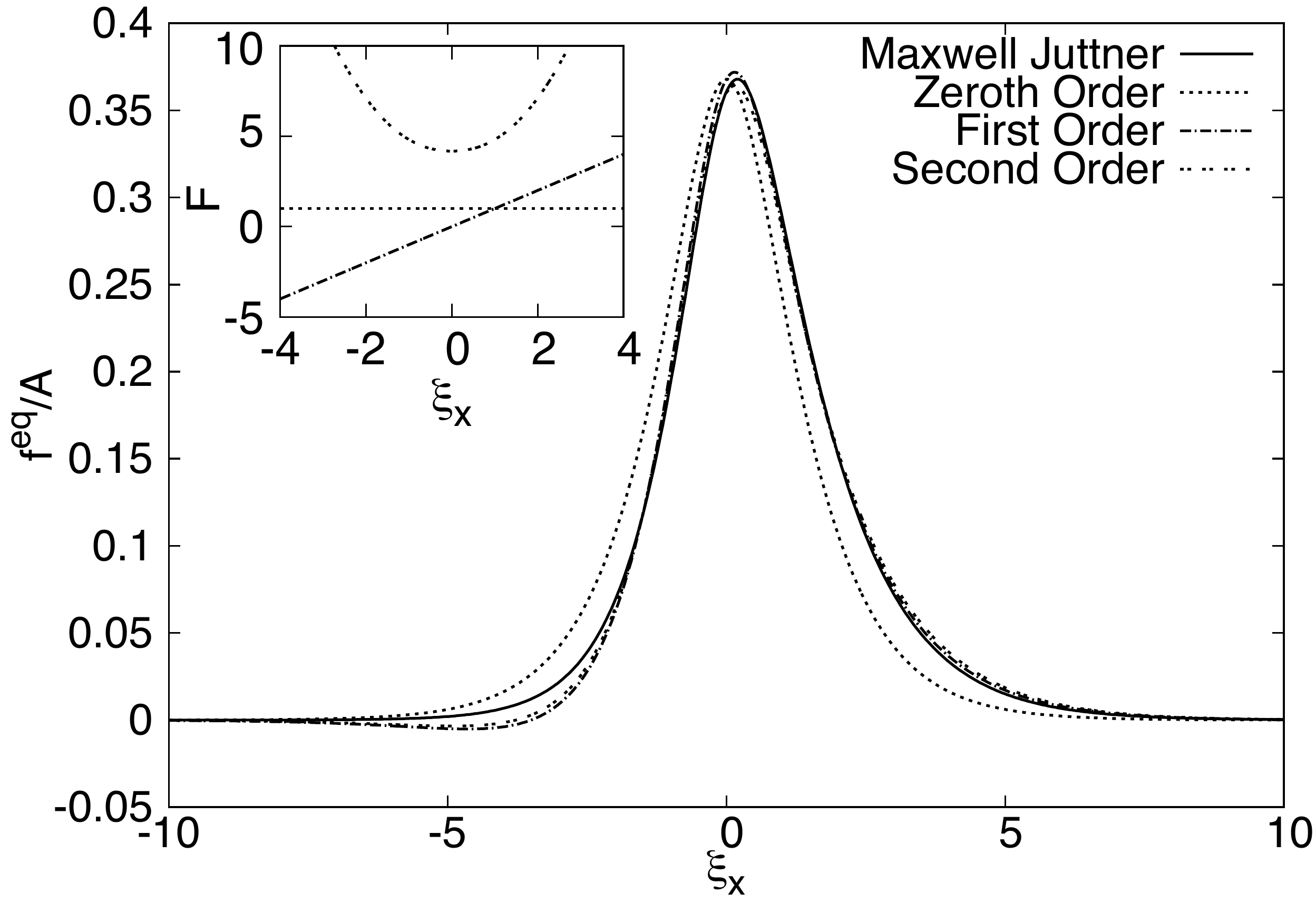}
\caption{Comparison between the Maxwell J\"uttner distribution
  function and the zeroth, first and second order expansions in one
  space dimension for $\beta=0.2$. In the inset, $F_{(0)}$},
$F_{(1)}^{x}$ and $F_{(2)}^{xx}$ ploynomials, in the local rest frame are shown.
\label{MJcomparison}
\end{center}
\end{figure}

One can compare the Maxwell J\"uttner distribution with the zeroth,
first and second order expansions in the one dimensional case. The
result of the distributions versus $\xi_x$ for the case
$\beta=|\boldsymbol{u}|/c=0.2$ is presented in
Fig. ~\ref{MJcomparison}. Here, we can observe that, as expected, as
the order of the expansion increases, the expansion becomes more
accurate.  Note that the expanded distributions become negative for
values of $\xi_x$, around $-3$. However, this is of no concern for our
model since, as we shall see shortly, the quadrature requires only
$\xi_x \sim 1$. 
For illustrative purposes, in the inset of
Fig.~\ref{MJcomparison}, we show, for $\nu
= 1$ and in the local rest frame, 
the polynomials corresponding to the zeroth ($F_{(o)}$), 
first ($F_{(1)}^x$), and second ($F_{(2)}^{xx}$) orders.

We can write the Boltzmann equation, Eq.\eqref{eqboltzmann}, as follows:
\begin{equation}\label{eqboltzmann2}
  \xi^0 \partial_t f+\xi^a \partial_a f = -\frac{\nu}{ \tau c}( f - f^{\rm eq} ),
\end{equation}
where latin subscript $a$ runs over the spatial coordinates. 
In order to discretize Eq.\eqref{eqboltzmann2} and avoid a 
multi-time lattice, we need to consider the temporal components of the discretized velocity
4-vector, i.e. $\xi^0_i$, as constant. 
Therefore, a transformation of the temporal component of 
both $\xi^\alpha$ and $\chi^\alpha$ is required. 
We can write
$(\xi_i^\alpha)=(c_t/c_0,\boldsymbol{c_a})$ and $(\chi^{\alpha})=(\chi^0/c_0, \boldsymbol{\chi}) $,
where $c_t$, $c_0$ and $c_a$ are constants related to the size of the
lattice. We will use a lattice configuration D3Q19 (19 discrete
vectors in 3 spatial dimensions), which can be expressed as
\begin{equation}
  \boldsymbol{c_a}=\left \{ \begin{array}{ll}
      (0,0,0) &  i=0; \\
      c_a(\pm 1,0,0)_{FS} &  1\leq i \leq 6; \\
      c_a(\pm 1,\pm 1,0)_{FS} &  7\leq i \leq 18, \end{array} \right.
\end{equation}
where the subscript $FS$ denotes a fully symmetric set of points.

To find the discretized weights for the lattice, we use a quadrature
procedure. According to the quadrature rule, the discretized weights
should satisfy the relation
\begin{equation}
  \int R(\boldsymbol{\xi})w(\boldsymbol{\xi})\,\frac{d^3\xi}{\xi^0} = \sum_{i=1}^N R(\xi_i)w_i ,
\end{equation}
where $R(\boldsymbol{\xi})$ is an arbitrary polynomial of order $2N$
or less. Using this relation, we can construct a system of equations
by replacing $R(\boldsymbol{\xi})$ with different combinations of
zeroth, first and second order polynomials. The left hand side of the
above equation can be calculated by using Eq.\eqref{moment1} to
\eqref{moment3}. Thus, the resulting discrete weights are given by
\begin{equation}
w_0 = 1+\frac{4 c_t^2 \nu^2}{361 c_0^2}-\frac{c_t^2}{c_a^2 c_0^2} ,
\end{equation}
\begin{equation}
w_i = \frac{c_t^2}{2166 c_0^2 c_a^2} \left(361-8 c_a^2 \nu^2\right) ,
\end{equation}
for $1\leq i \leq 6$, and
\begin{equation}
w_i = \frac{c_t^2 \nu^2}{1083 c_0^2},
\end{equation}
for $7\leq i \leq 18$. Note that we still need to calculate the
constants related to the size of the lattice, i.e.  $c_a, c_t$ and
$c_0$.

The discretized 4-momenta should satisfy the following relation for
the energy-momentum tensor, i.e. the second order moment of the
distribution function,
\begin{eqnarray}\label{energymomentum}
\int p^\alpha p^\beta f^{\rm eq} \,\frac{d^3p}{p^0} &=&  \sum_{i=1}^N p_i^\alpha p_i^\beta f_i^{\rm eq} \nonumber\\
&=& (\epsilon+p) \frac{U^\alpha U^\beta}{c^2} - p \eta^{\alpha\beta}=T_{\rm eq}^{\alpha \beta} ,
\end{eqnarray}
where $p$ is the hydrostatic pressure, $\epsilon$ the energy density
and $T_{\rm eq}^{\alpha \beta}$ denotes the energy-momentum tensor at
equilibrium. Note that higher order moments of the
  discrete equilibrium distribution can be calculated by performing
  the respective sums $T^{\alpha \beta ... \gamma} = \sum_{i=1}^N
  p_i^\alpha p_i^\beta ...p_i^\gamma f_i^{\rm eq}$. However, they would
  not correspond to the ones of the Maxwell-J\"uttner distribution,
  because the latter require an expansion in higher order
  polynomials. Their contribution to the dynamics of the fluid become
  important at high values of the Knudsen number (high momentum diffusivity),
  and since they are not exactly recovered, our model does not work
  properly in that regime. Fortunately, many applications in
  astrophysics and high energy physics deal with near-inviscid or weakly
  viscous fluids.

We can simply find the constants related to the lattice
size, using the fact that in the tensor, the coefficient of $U^\alpha
U^\beta$ for different $\alpha$ and $\beta$ should be always the
same. The calculated values for the constants are
\begin{equation}
c_a=\frac{\sqrt{19}}{\nu}, \quad c_t/c_0=\frac{\sqrt{27}}{\nu}, \quad c_0=\frac{3}{8}(9-2\sqrt{3}).
\end{equation}
 
In the ultrarelativistic limit and considering the natural units
$c=k_B=1$, from the energy-momentum tensor, one can obtain the
following relations:
\begin{equation}
  \epsilon+p =  \frac{4n}{\nu^2} , \quad p =  \frac{n}{\nu^2}, \quad
  \epsilon = 3p \quad ,
\end{equation}
finding that the relation between $\epsilon$ and $p$ corresponds to
the well-known state equation in the ultrarelativistic limit.

Note that due to the fact that we have supposed $\xi^0_i$ to be
constant, to avoid a multi-time evolution lattice, there are some
equations in the quadrature procedure for the first order moment and
the second order moment of the distribution function which could not
be satisfied simultaneously. Indeed, we can choose whether to recover
the first order moment or the second order moment in the
quadrature. To satisfy the first moment of the Maxwell-J\"uttner
distribution function leads to recover the equation for the
conservation of number of particles, $\partial_\alpha N^\alpha = 0$,
and the second order moment, the equation for the conservation of
momentum-energy, $\partial_\alpha T^{\alpha \beta} = 0$. To calculate the four unknowns, namely $U^x, U^y, U^z$ and $\epsilon$, the four equations corresponding to $T^{00}, T^{0x}, T^{0y}$ and $T^{0z}$ components of
the energy-momentum tensor and equation of state $\epsilon = 3p$ would be enough. Therefore, by using the ultrarelativistic equation of state the dynamics of the system
is not affected by the number of particles and the equation for
the conservation of momentum-energy is therefore sufficient to describe the
entire dynamics of the relativistic fluid. For this reason, our quadrature is targeted to recover the second order moment, 
using a separate distribution function to recover the equation for the
conservation of number of particles, $\partial_\alpha N^\alpha = 0$,
based on the model proposed by Hupp et al. \cite{hupp2011relativistic}.

Using the mentioned lattice to discretize the Boltzmann equation,
Eq.\eqref{eqboltzmann2} can be written as follows:
\begin{equation}
f_i(\boldsymbol{x}+\boldsymbol{c_a}\frac{c_0}{c_t} \delta t, t+ \delta t ) - f_i \left(\boldsymbol{x}, t \right) = - \frac{c_0 \nu \delta t}{\tau c_t}\left(f_i -f_i^{\rm eq} \right).
\end{equation}
The left hand side of the equation is readily recognized as
free-streaming, while the right hand side is the discrete version of the
collision operator according to the model of Marle. 
In this equation, the following relation between $\delta t$ and 
$\delta x$ has been used:
\begin{equation}\label{deltat}
\delta t = \frac{c_t \delta x}{c_a c_0}.
\end{equation}

In the ultra-relativistic limit, the shear viscosity using the model of
Marle can be calculated as:
\begin{equation}\label{viscosity}
\eta = \frac{(\epsilon+p)}{\nu^2} \left (\tau -\frac{1}{2} \right ).
\end{equation}

In our model, this expression is only valid for small
  values of $\tau$ (which leads to small values of the Knudsen
  number), where higher order moments of the distribution can be
  neglected.

At each time step and at each lattice point, the values of the
macroscopic velocity and energy density can be evaluated using the
energy-momentum tensor as mentioned previously. 

\subsection{Extended model for high velocities}

At high velocities (${\beta} > 0.6$), due to the compressibility
effects (high Mach numbers), the described numerical scheme shows
artificial discontinuities in the velocity and pressure profiles,
leading to numerical instabilities in the long-term evolution.  
We shall return to this issue in Sec.~\ref{Validation and Results}. 
The relativistic Mach number can be expressed as
$M^R=\gamma(u)|\boldsymbol{u}|/\gamma(c_{so}) c_{so}$ where $c_{so}$
is the velocity of sound, which is $c_{so}=c/\sqrt{3}$ in the
ultra-relativistic regime. In order to overcome this problem, we first
use a modified version of the D3Q19 cell configuration, which is
denoted by $(\xi_i^\alpha)=(c_t/c_0,\boldsymbol{c_a})$, where
$\boldsymbol{c}_a$ takes now the following form:
\begin{equation}
\boldsymbol{c_a}=\left \{ \begin{array}{ll}
(0,0,0) &  i=0; \\
c_a(\pm 1,0,0)_{FS} &  1\leq i \leq 6; \\
2 c_a(\pm 1,\pm 1,0)_{FS} &  7\leq i \leq 18. \end{array} \right.
\end{equation}
Since some of the discrete velocities go beyond first and second
neighbors in the lattice, the scheme can support higher
flow speeds. In order to find the discretized weights, as well as the
size of the lattice cells, we repeat the same procedure as before
obtaining
\begin{equation}
w_0 = 1+\frac{7 c_t^2 \nu^2}{676 c_0^2}-\frac{c_t^2}{c_a^2 c_0^2},
\end{equation}
\begin{equation}
w_i = \frac{c_t^2}{1014 c_0^2 c_a^2} \left(169-2 c_a^2 \nu^2\right),
\end{equation}
for $1\leq i \leq6$, 
\begin{equation}
w_i = \frac{c_t^2 \nu^2}{8112 c_0^2},
\end{equation}
for $7\leq i \leq 18$, and
\begin{equation}
c_a=\frac{\sqrt{13}}{\nu}, \quad c_t/c_0=\frac{\sqrt{27}}{\nu}, \quad c_0=\frac{3}{8}(9-2\sqrt{3}).
\end{equation}

The second step to extend our model, is to introduce the minimum
modulus (min mod) scheme to discretize the spatial components in the
streaming term in the Boltzmann equation, Eq.\eqref{eqboltzmann2},
i.e. $p^a_i \partial _a f_i$. The min mod scheme is a flux limiter
method that efficiently reduces the instability especially when step
discontinuities occur (e.g. in shock waves). The following relations
characterize this scheme \cite{pan2007lattice}:
\begin{equation}
\partial _a (p^a_i f_i) = \frac{1}{|\delta x \boldsymbol{e}_a|} \left[ h_i^a(\boldsymbol{x}+\delta x  \boldsymbol{e}_a) - h_i^a(\boldsymbol{x}) \right],
\end{equation}
\begin{equation}
h_i^a(\boldsymbol{x} ) = f_i^{a^L}(\boldsymbol{x} )+ f_i^{a^R}(\boldsymbol{x}), 
\end{equation}
\begin{equation}
\begin{array}{c}
f_i^{a^L}(\boldsymbol{x}) = f_i^{a^+}(\boldsymbol{x} ) \\
+ \frac{1}{2} \mbox{min mod}\left(\triangle f_i^{a^+}(\boldsymbol{x} ), \triangle f_i^{a^+}(\boldsymbol{x}-\delta x  \boldsymbol{e}_a )\right),
\end{array}
\end{equation}
\begin{equation}
\begin{array}{c}
f_i^{a^R}(\boldsymbol{x} ) = f_i^{a^-}(\boldsymbol{x}+\delta x  \boldsymbol{e}_a) \\
- \frac{1}{2} \mbox{min mod}\left(\triangle f_i^{a^+}(\boldsymbol{x} ), \triangle f_i^{a^+}(\boldsymbol{x}+\delta x  \boldsymbol{e}_a )\right),
\end{array}
\end{equation}
\begin{equation}
f_i^{a^+} = \frac{1}{2}(p_i^a+|p_i^a|)f_i, \quad f_i^{a^-} = \frac{1}{2}(p_i^a-|p_i^a|)f_i,
\end{equation}
\begin{equation}
\triangle f_i^{a^\pm}(\boldsymbol{x} ) = f_i^{a^\pm}(\boldsymbol{x}+\delta x  \boldsymbol{e}_a) - f_i^{a^\pm}(\boldsymbol{x}), 
\end{equation}
where $\boldsymbol{e}_a$ is a unit vector in the direction of the
corresponding spatial coordinate. Let us remind that $p^a_i$ is
independent of spatial coordinates. The min mod function is defined as
\begin{equation}
\begin{array}{c}
\mbox{min mod}(X,Y) = \frac{1}{2} \mbox{min} (|X|,|Y|) \\ 
\times [\mbox{Sign}(X)+\mbox{Sign}(Y)].
\end{array}
\end{equation}

Note that in classical flows, the bulk viscosity plays an important
role in highly compressible flows and enhances the stability of
numerical simulations of fluids at very high velocities, including
shock waves \cite{dellar2001bulk}. However, in the ultrarelativistic
limit, the energy-momentum tensor is traceless and the bulk viscosity
is zero \cite{romatschke2010new,
mendoza2013ultrarelativistic}. 
Therefore, in order to include the bulk viscosity, we add the
following term to the right hand side of the Boltzmann
equation, Eq.\eqref{eqboltzmann2}: 
\begin{equation}
\lambda_i \sum_{a=x,y,z} \partial_a^2 f_i ,
\end{equation}
where
\begin{equation}
 \lambda_i = \left\{ \begin{array}{ll}
0 & i=0 ; \\
 \alpha \delta x & i\not= 0,\end{array} \right. 
\end{equation}
where $\alpha$ is a constant. A central finite difference scheme is
used to calculate the second order derivative. 
Chapman-Enskog analysis reveals that the bulk viscosity 
obtained by this extra-term takes the form
\begin{equation}
\eta_b =  \frac{4 p \alpha \nu^6}{27}.
\end{equation}
Note that, like the analytical expression of the bulk viscosity for the
model of Marle, the above-mentioned bulk viscosity is also
proportional to $T^{-3}$, and as expected goes to zero in the
ultra-relativistic limit $\nu \rightarrow 0$. 
However, this small value of bulk viscosity is sufficient to stabilize the system at
high velocities, as we are going to show in the next section.

Once all of the extensions above are taken into account, the
discretized form of the relativistic Boltzmann equation 
takes the following expression:
\begin{equation}\label{eq:rlb}
\begin{array}{c}
f_i(\boldsymbol{x}, t+\delta t) - f_i(\boldsymbol{x}, t) \\ 
+ \frac{c_0}{c_t} \frac{\delta t}{\delta x} \left[ h_i^a(\boldsymbol{x}+\delta x  \boldsymbol{e}_a) - h_i^a(\boldsymbol{x}) \right] = \\ 
-\frac{c_0 \nu \delta t}{\tau c_t}\left[f_i(\boldsymbol{x},t+\delta t) - (2f_i^{\rm eq}(\boldsymbol{x},t)-f_i^{\rm eq}(\boldsymbol{x},t-\delta t)) \right] \\
+\frac{c_o \nu \delta t}{c_t} \lambda_i \sum \partial_a^2 f_i(\boldsymbol{x},t),
\end{array}
\end{equation}
where an implicit representation of the collision term is used, as
proposed in Ref. \cite{mei1998finite} to enhance the stability
of the collision term.

As mentioned above, for the cases where the dynamics of the number of
particles density is also needed, one has to solve the conservation
equation, i.e. $\partial_\alpha N^\alpha=0$, with
$N^\alpha=nU^\alpha$. For this purpose, we add an extra
distribution function, $h_i$, based on the model proposed by Hupp et
al. \cite{hupp2011relativistic}, which follows the dynamics of the Boltzmann
equation given by Eq.~\eqref{eq:rlb}, without the $\lambda_i$
coefficient term. 
The corresponding modified equilibrium distribution function is given by:
\begin{equation}\label{Hupp}
h_i^{\rm eq}=w'_i n \gamma(u)\left(\frac{c_0}{c_t}+3(\boldsymbol{c_a}.\boldsymbol{u})+\frac{9}{2}(\boldsymbol{c_a}.\boldsymbol{u})^2-\frac{3}{2}|\boldsymbol{u}|^2\right),
\end{equation}
\begin{equation}
w'_0=\frac{1}{10}, \quad w'_i=\frac{6}{35}-\frac{1}{42 c_a^2},
\end{equation}
for $1\leq i \leq6$, 
\begin{equation}
w'_i=\frac{1}{84 c_a^2}-\frac{3}{280}
\end{equation}
for $7 \leq i \leq 18$, where we have used our new cell configuration,
and $w'_i$ are the respective discrete weights.

Having discussed the model, we move on to the next
section, where different validations and results for the Riemann
problem are provided, along with a simulation of a shock wave
colliding with a interstellar cloud.

\section {Validation and Results}
\label {Validation and Results} 

In order to validate the model and the numerical procedure, we present
results for the simulation of relativistic shock wave propagation in 
viscous quark-gluon plasma.
The associated Riemann problem is studied and several comparisons are
drawn between the present RLB model and the existing literature. 
Indeed, the Riemann problem is a challenging test for numerical methods, since 
it involves a shock and rarefaction wave.

The initial condition of the Riemann problem consists of two regions
with different pressure, which are separated by a membrane in the
middle of the interval. The pressure in the left region ($P_0$) is
higher than the pressure in the right region ($P_1$). Both sides of
the discontinuity are supposed to be initially in the rest frame.
Hence, spatial components of initial velocities for both sides
are set to zero. At time $t=0$, the membrane is removed and a shock
wave propagates from the high pressure region into the low pressure
region with velocity $v_{shock}$ and a rarefaction wave propagates in
the opposite direction. The shock velocity only depends on the
pressure difference, the equation of state, and can be calculated
analytically \cite{Rischke1995346,thompson1986special}. The region
between the shock wave and the rarefaction wave has a constant
pressure, corresponding to the so-called shock plateau. 
In this region the velocity is also constant ($v_{plat}$).

\begin{figure}
\begin{center}
\includegraphics[trim=0mm 0mm 0mm 0mm, clip, width=0.86\columnwidth,, height=0.6\columnwidth ]{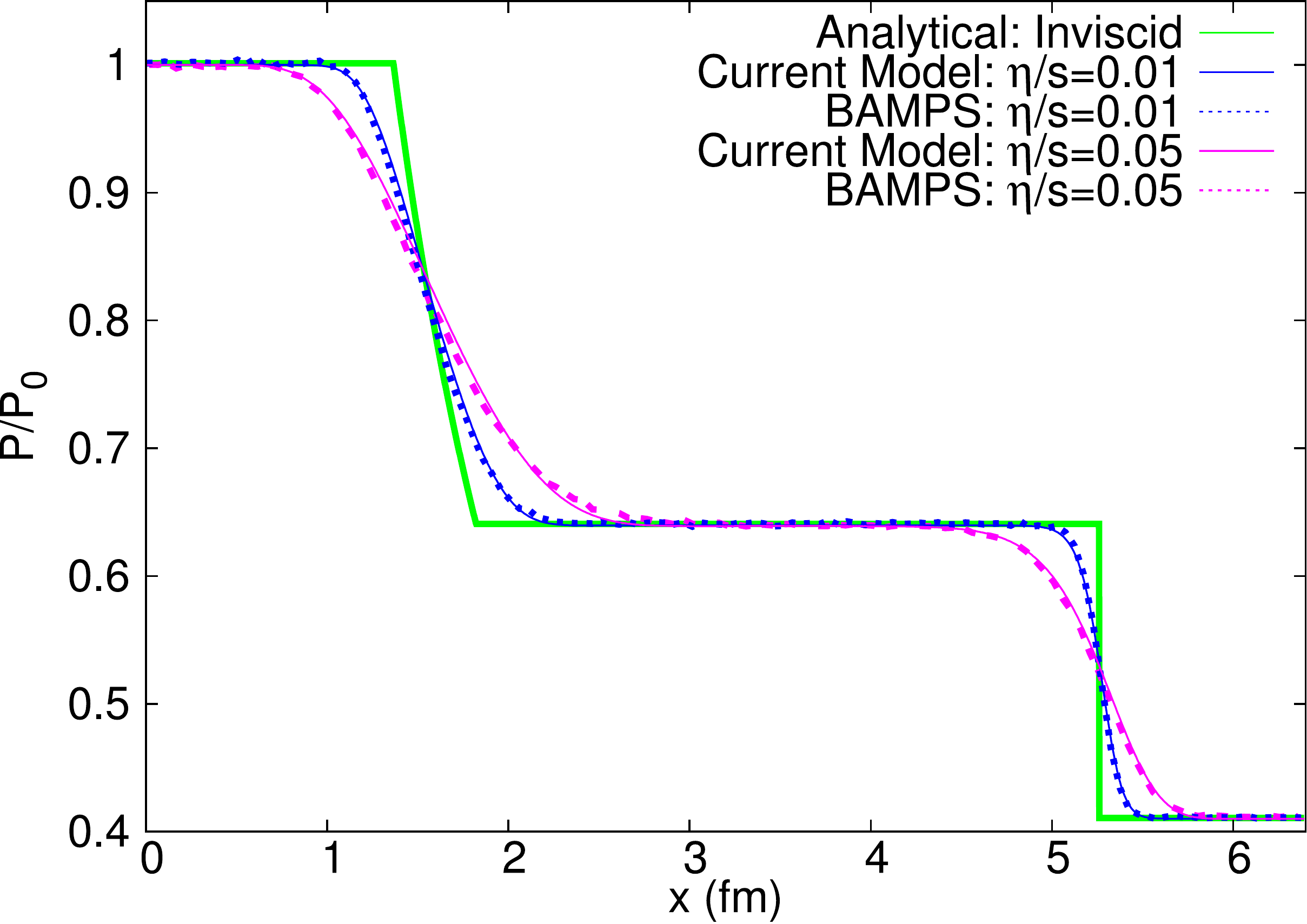}
\caption{Comparison of the velocity profile of the current model
and BAMPS, for different values of $\eta/s$ at weakly relativistic regime.}
\label{comparebampsmillerme2}
\end{center}
\end{figure}

\begin{figure}[h]
\begin{center}
\includegraphics[trim=10mm 20mm 10mm 10mm, clip, width=0.9\columnwidth, height=0.6\columnwidth ]{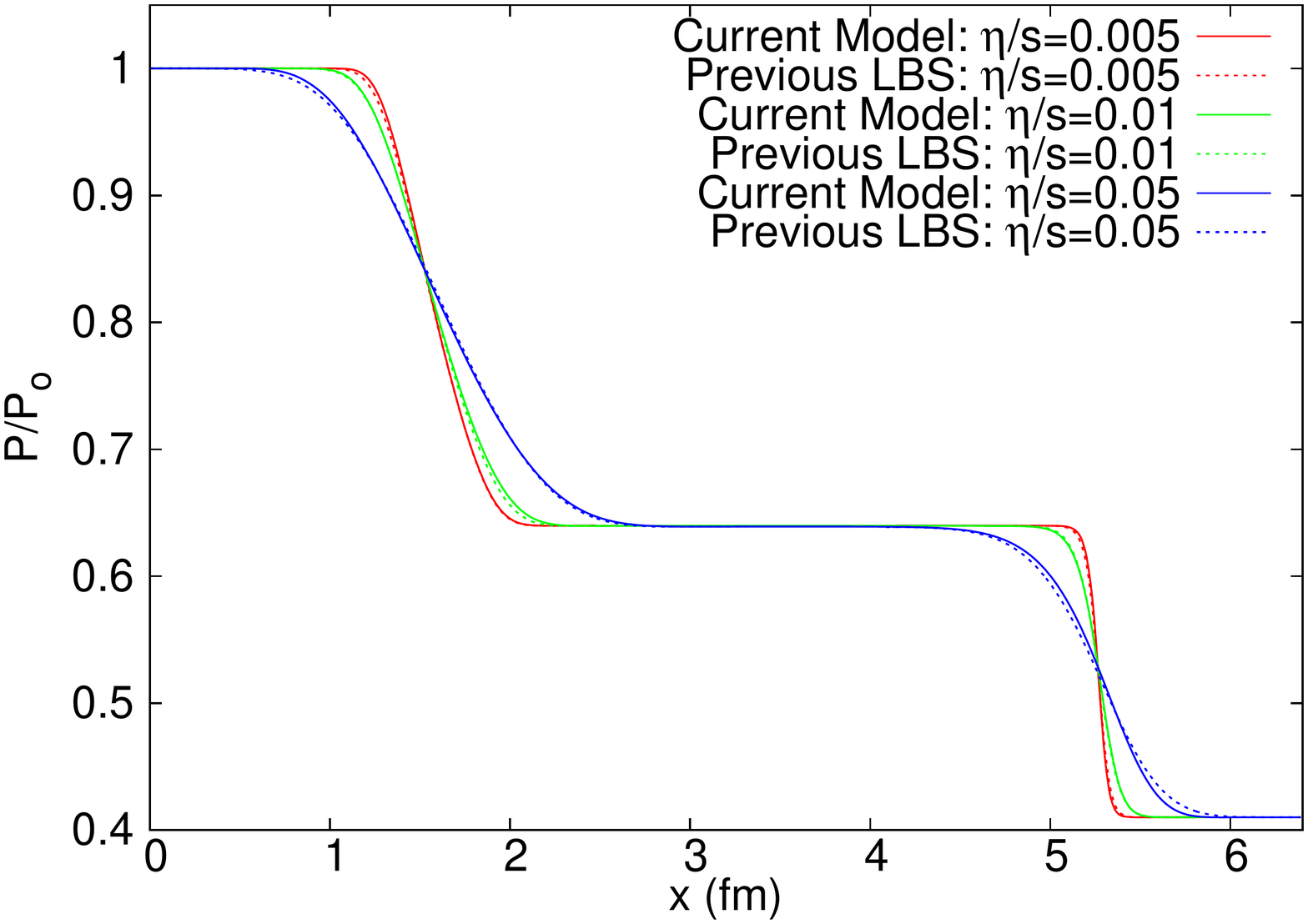}
\includegraphics[trim=10mm 20mm 10mm 10mm, clip, width=0.9\columnwidth, height=0.6\columnwidth ]{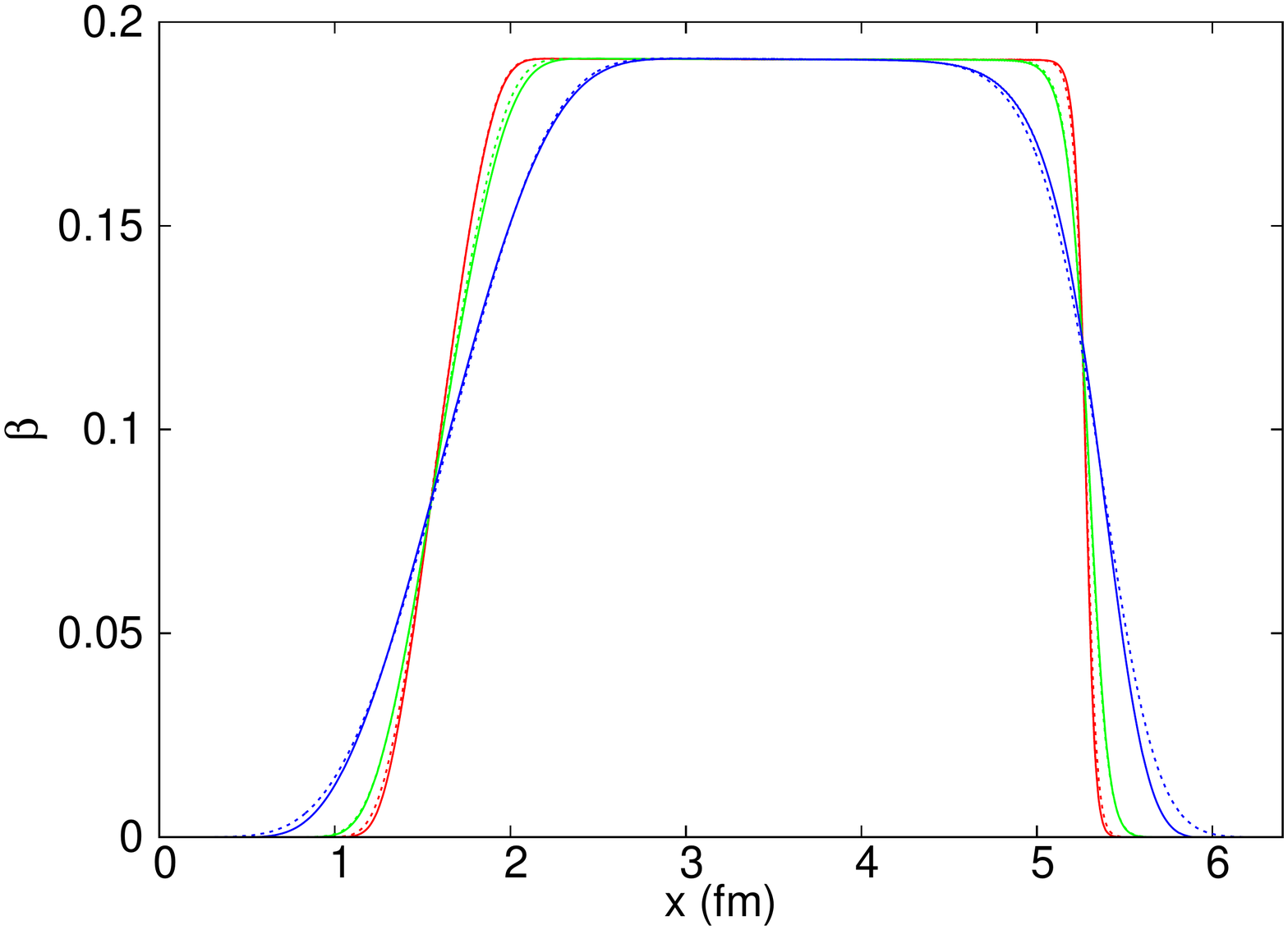}
\end{center}
\caption{Comparison between the current model and the previous LBS
  model at different $\eta/s$, for the pressure (top) and velocity
  (bottom) profiles in the weakly relativistic regime.}
\label{millermecompare2eta}
\end{figure}

\begin{figure}[h]
\begin{center}
\includegraphics[trim=10mm 20mm 10mm 10mm, clip, width=0.9\columnwidth, height=0.6\columnwidth ]{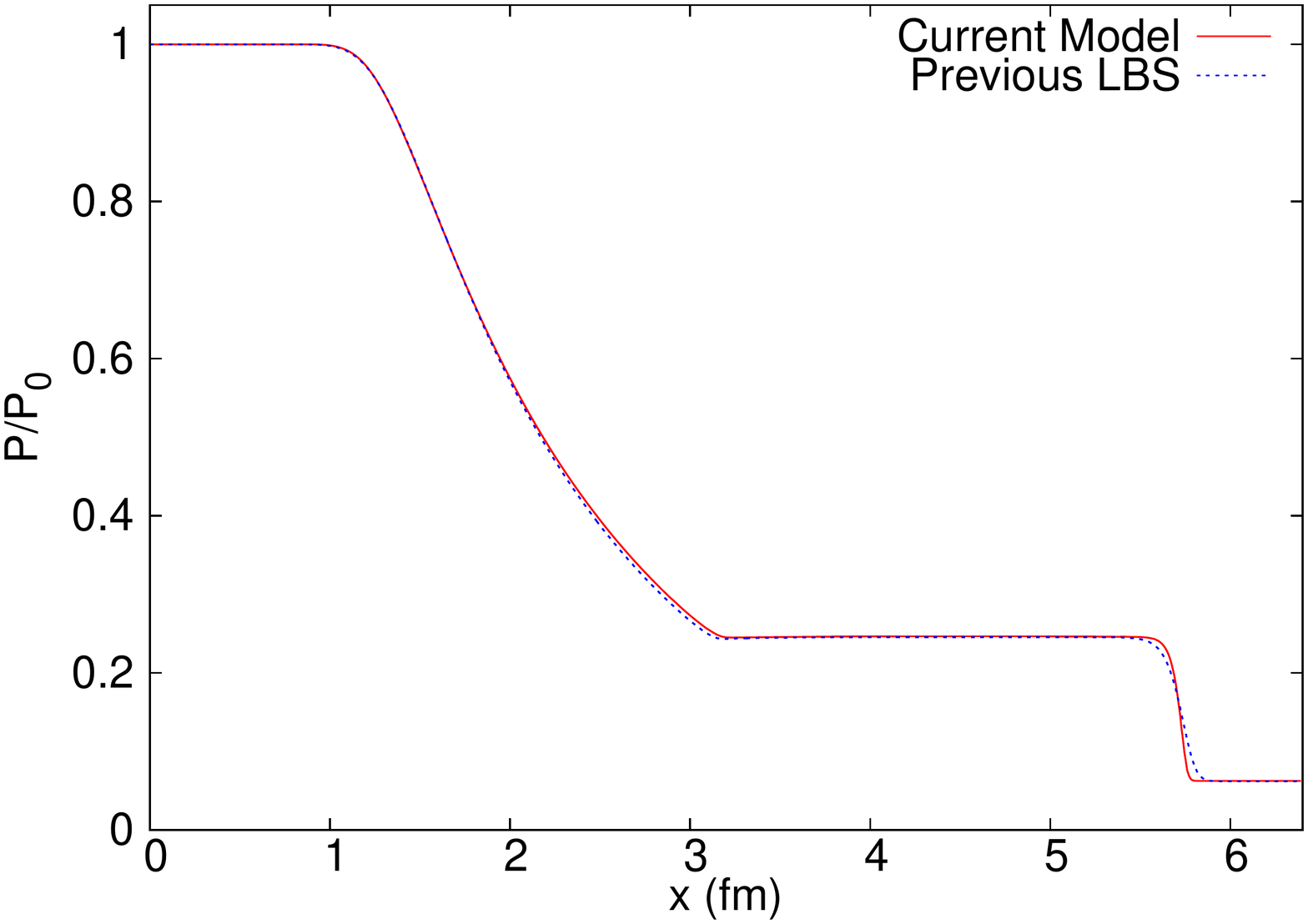}
\includegraphics[trim=10mm 20mm 10mm 10mm, clip, width=0.9\columnwidth, height=0.6\columnwidth ]{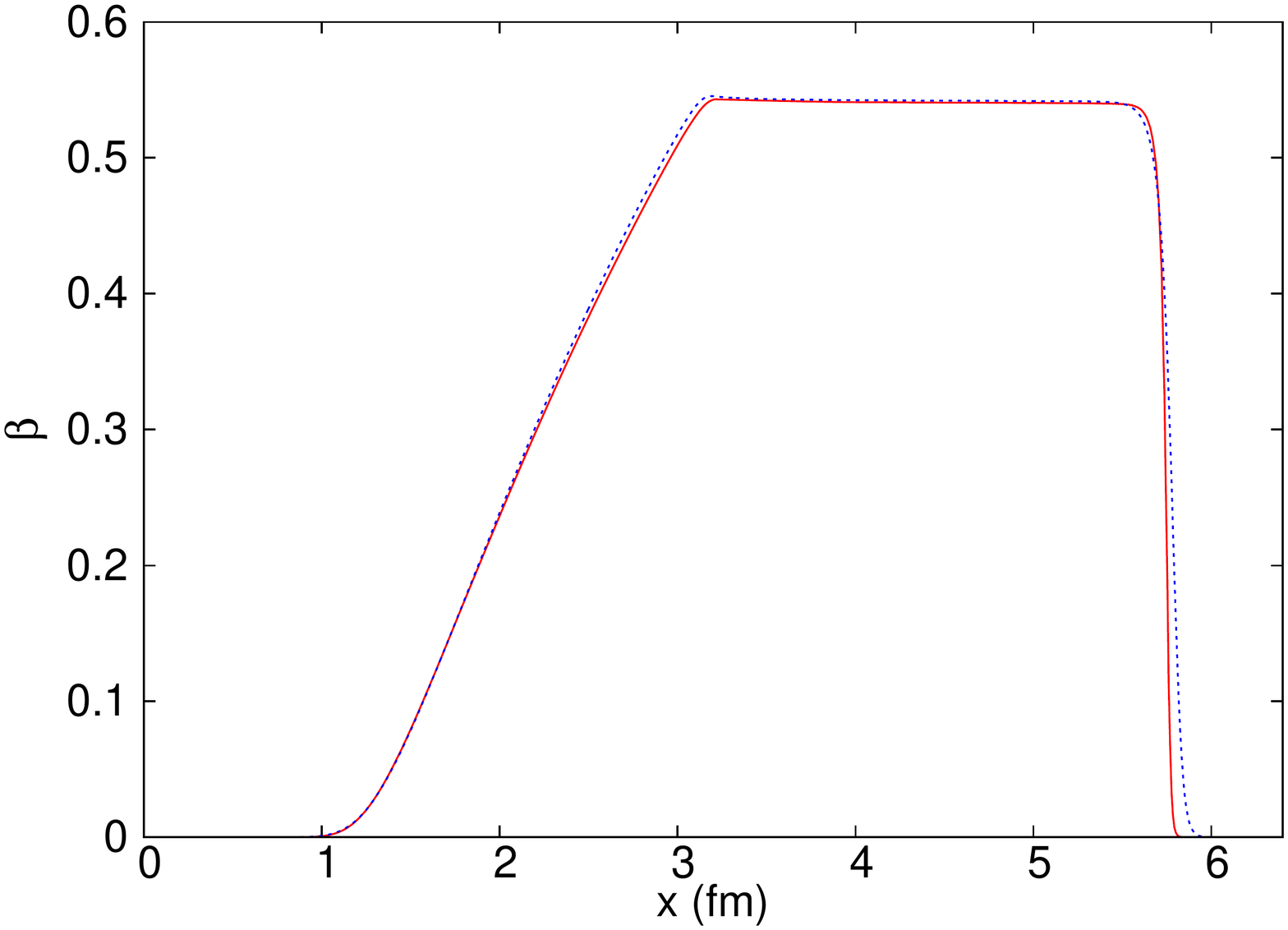}
\end{center}
\caption{Comparison between the current model and the previous LBS
  model at $\eta/s=0.01$, for the pressure (top) and velocity (bottom)
  profiles in the moderately relativistic regime.}
\label{millercompare6}
\end{figure}

\begin{figure}[h]
\begin{center}
\includegraphics[trim=10mm 20mm 10mm 10mm, clip, width=0.9\columnwidth, height=0.6\columnwidth ]{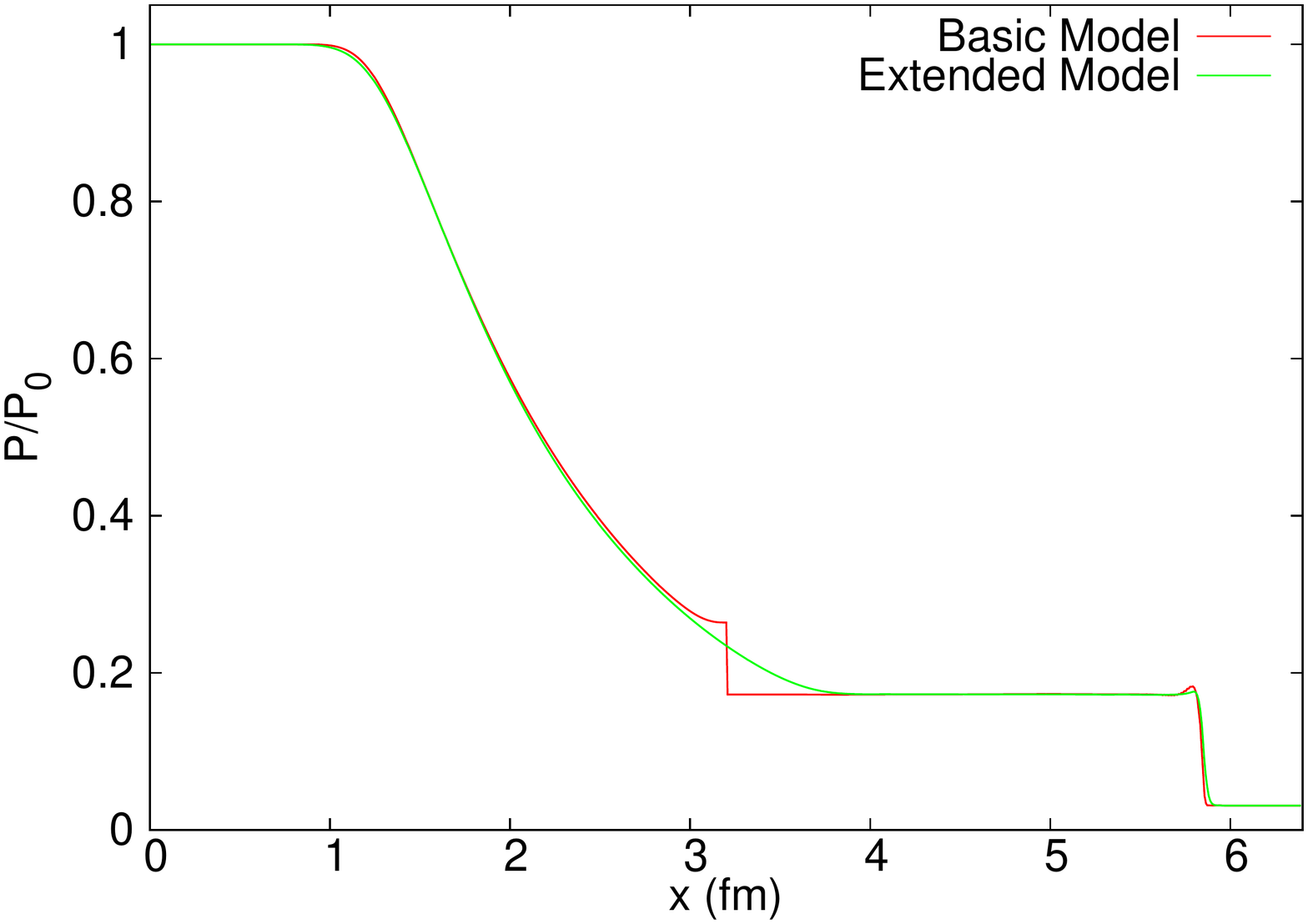}
\includegraphics[trim=10mm 20mm 10mm 10mm, clip, width=0.9\columnwidth, height=0.6\columnwidth ]{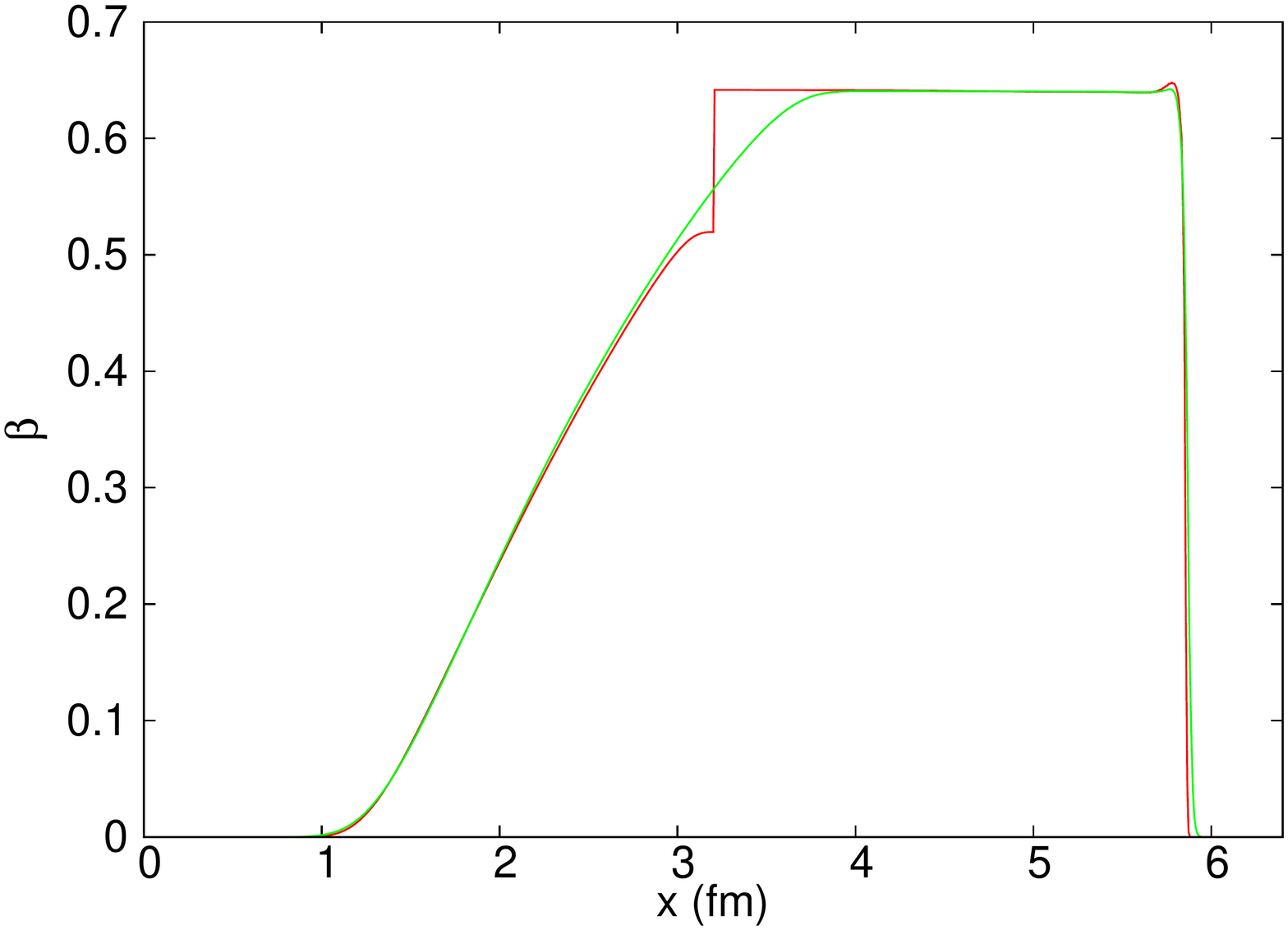}
\end{center}
\caption{Comparison between the basic and extended models for the
  pressure (top) and velocity (bottom) profiles in the moderately relativistic regime. }
\label{basicextended}
\end{figure}

In order to compare our results with existing models, we use
the same conditions as in
Ref. \cite{bouras2009relativistic,PhysRevLett.105.014502}. 
Therefore, one dimensional simulations are carried out using $800\times1\times1$
cells, where open boundary conditions are considered at the two
ends. The cell size $\delta_x$ is taken to be unity, which
corresponds to $\delta_x=0.008 fm$ in IS units and $\delta_t$ can be
calculated from Eq.\eqref{deltat}. We use $\eta/s$ as the
characteristic parameter of shear viscosity, where the viscosity is
defined in Eq.\eqref{viscosity} and the entropy density is 
given by $s=4n-n \ln \lambda$, with $\lambda=n/n^{eq}$ being the fugacity of gluons,
$n^{eq}=d_G T^3/\pi^2$ the equilibrium density, and $d_G=16$ the
degeneracy of gluon.

For the first validation test, we set the initial pressure at the left
and right sides to $P_0=5.43 GeV /fm^3$ and $P_1=2.22 GeV /fm^3$,
respectively. This corresponds to $2.495\times10^{-7}$ and
$1.023\times10^{-7}$ in numerical units, respectively. 
Fig.~\ref{comparebampsmillerme2} shows the pressure
profiles at time $t=3.2 fm/s$ for different values of $\eta/s$,
compared to the results reported by Ref. \cite{bouras2009relativistic}
(hereafter BAMPS) and an analytical solution for the inviscid case
reported in Ref. \cite{Rischke1995346} . As one can notice, very
satisfactory agreement for different values of $\eta/s$ is obtained..

As mentioned previously, the lattice Boltzmann method is 
computationally very efficient. 
For instance, the above simulation took $\sim 220$ ms on a
standard PC, which is approximately an order of magnitude faster than
corresponding hydrodynamic simulations. To further elaborate on the
validity of our model, we compare with the results of
Ref. \cite{PhysRevLett.105.014502} (hereafter previous LBS) for
different values of $\eta/s$. 
Fig.~\ref{millermecompare2eta} shows
that pressure and velocity profiles are in good agreement
with previous LBS simulations. 
It is worth mentioning that, as it is  apparent from Fig.~\ref{millermecompare2eta},
the above mentioned pressure difference corresponds to
$\beta=|v_{plat}|\sim0.2$ (weakly relativistic regime), while the
velocity of the shock is $v_{shock}\sim0.65$.

To study higher velocities, we consider higher pressure difference
between the left side and the right side, namely $P_0=5.43GeV /fm^3$
and $P_1=0.339GeV /fm^3$. This corresponds to $ 2.495\times10^{-7}$
and $ 1.557\times10^{-8}$ in numerical units, respectively. The
resulting pressure and velocity profiles for $\eta/s=0.01$, are
compared to the results with previous LBS in Fig.~\ref{millercompare6},
showing again very good agreement. 
It should be noted that, due to the above-mentioned pressure difference, the 
matter behind the shock moves with the velocity $\beta \sim 0.6$ 
(moderately relativistic regime), while the shock itself goes with the 
velocity $v_{shock} \sim 0.92$.

Note that the proposed model in the non-extended form (hereafter basic
model) becomes numerically unstable for higher velocities ($\beta>
0.6$). This instability is due to compressibility effects. 
It is known that the lattice Boltzmann method is intrinsically suited to low Mach
number flows (low compressibility effects). Therefore, in order to overcome
this problem, we use our extended model, which enhances
the stability of the numerical procedure without any appreciable loss
of computational efficiency. 
To further investigate this issue, we carry out two simulations with the same conditions for
relatively higher velocity, one using the basic model, and the other
one using the extended one. The pressure is set to be $P_0=5.43 GeV
/fm^3$ and $P_1=0.1695 GeV/fm^3$ for the left side and the right side,
respectively, which corresponds to $ 2.495\times10^{-7}$ and $
7.785\times10^{-9}$ in numerical units. Here, $\eta / s=0.01$, 
$\delta_t/\delta_x=0.25$, and $ \alpha=0.15$ for the extended model.

The results for the pressure and velocity profiles are shown in
Fig.~\ref{basicextended} at time $t=3.2 fm/s$. Note that the applied
pressure difference corresponds to $\beta \sim 0.7$. Using the basic
model, an artificial discontinuity is observed in both the pressure
and velocity profiles, which is due to instability problems in the
numerical scheme. However, the extended model proves capable
of handling the simulation, the problem of the artificial
discontinuity being solved completely. Additionally, apart from the region
affected by the discontinuity, the good agreement
between the results of two models can be interpreted as a validation
for the precision of the extended model. 
The required CPU time for the simulation using the extended model and
for the chosen value of $\delta_t/\delta_x$ is $1069$ ms.

To the best of our knowledge, to date, there was no reported
simulation of shock wave in viscous flow for $\beta>0.6$. 
However, for the inviscid case, which corresponds to the Euler equation at
macroscopic scale, there exists an analytical solution for the Riemann
problem. Therefore, in order to validate our extended model, we
compare our results with the analytical solution of the inviscid case
in Ref. \cite{thompson1986special}. Hence, we need to solve the Euler
equation by ignoring the viscous effects. It is worth mentioning that,
in the classical lattice Boltzmann method, the numerical solution
becomes unstable as one tries to set the shear viscosity to zero
($\tau = 1/2$). This is also the case for our basic model. However, in
the extended model we can solve the Euler equation by changing the
collision step, such that instead of implementing the regular
collision, we simply set the discretized distribution functions to
their corresponding equilibrium values. This is similar to the procedure used in Ref.\cite{nadiga1995euler} to solve Euler equation in the non-relativistic case. As one can notice from the Boltzmann equation, this corresponds to ignore the non-equilibrium
part of the distribution function which contains the information about
the dissipation. Therefore, we neglect the viscous effects in the
macroscopic equations, obtaining the Euler equations.

\begin{figure}[h]
\begin{center}
\includegraphics[trim=0mm 0mm 0mm -8mm, clip, width=0.9\columnwidth,, height=0.6\columnwidth ]{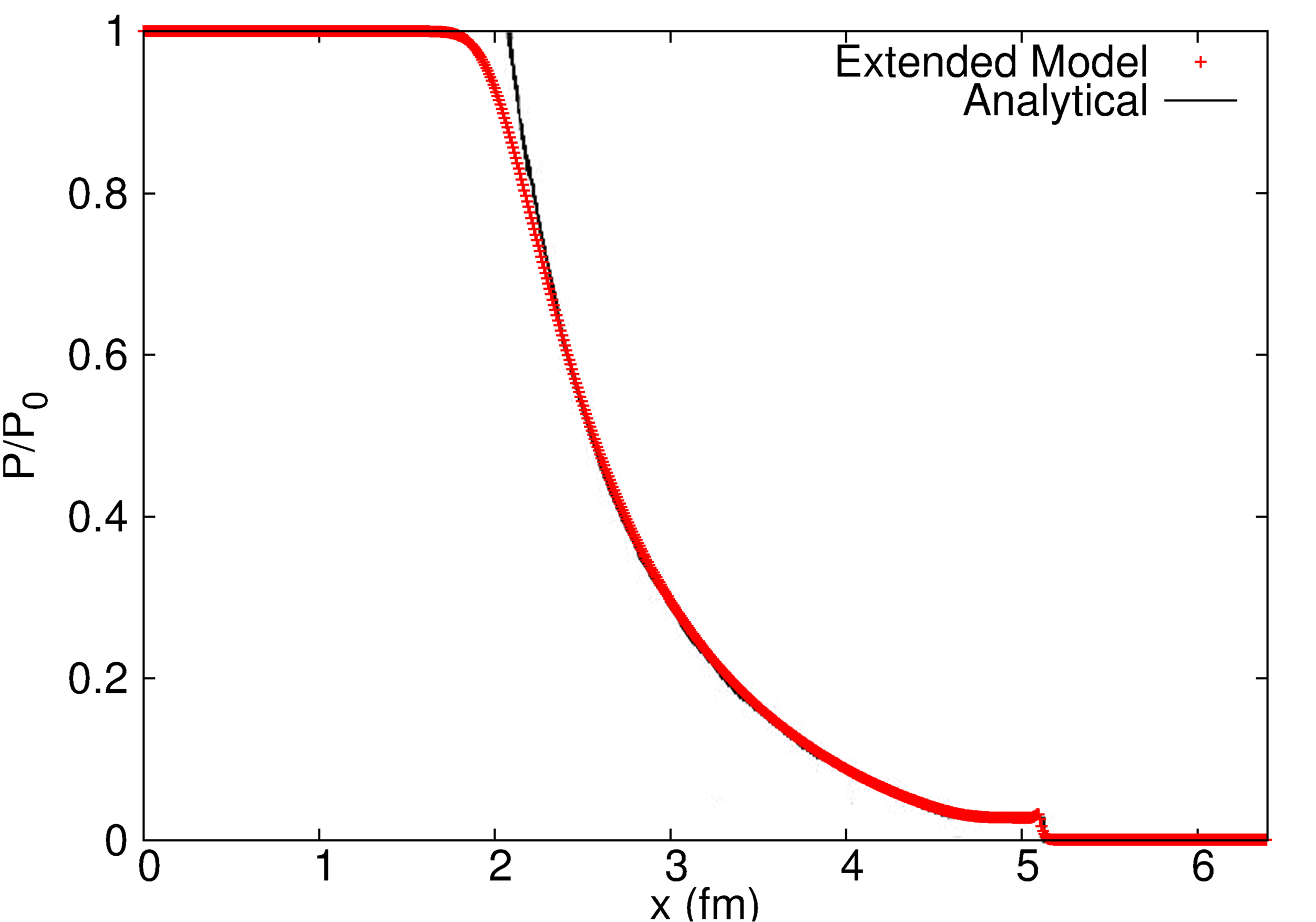}
\includegraphics[trim=0mm 0mm 0mm -8mm, clip, width=0.9\columnwidth,, height=0.6\columnwidth ]{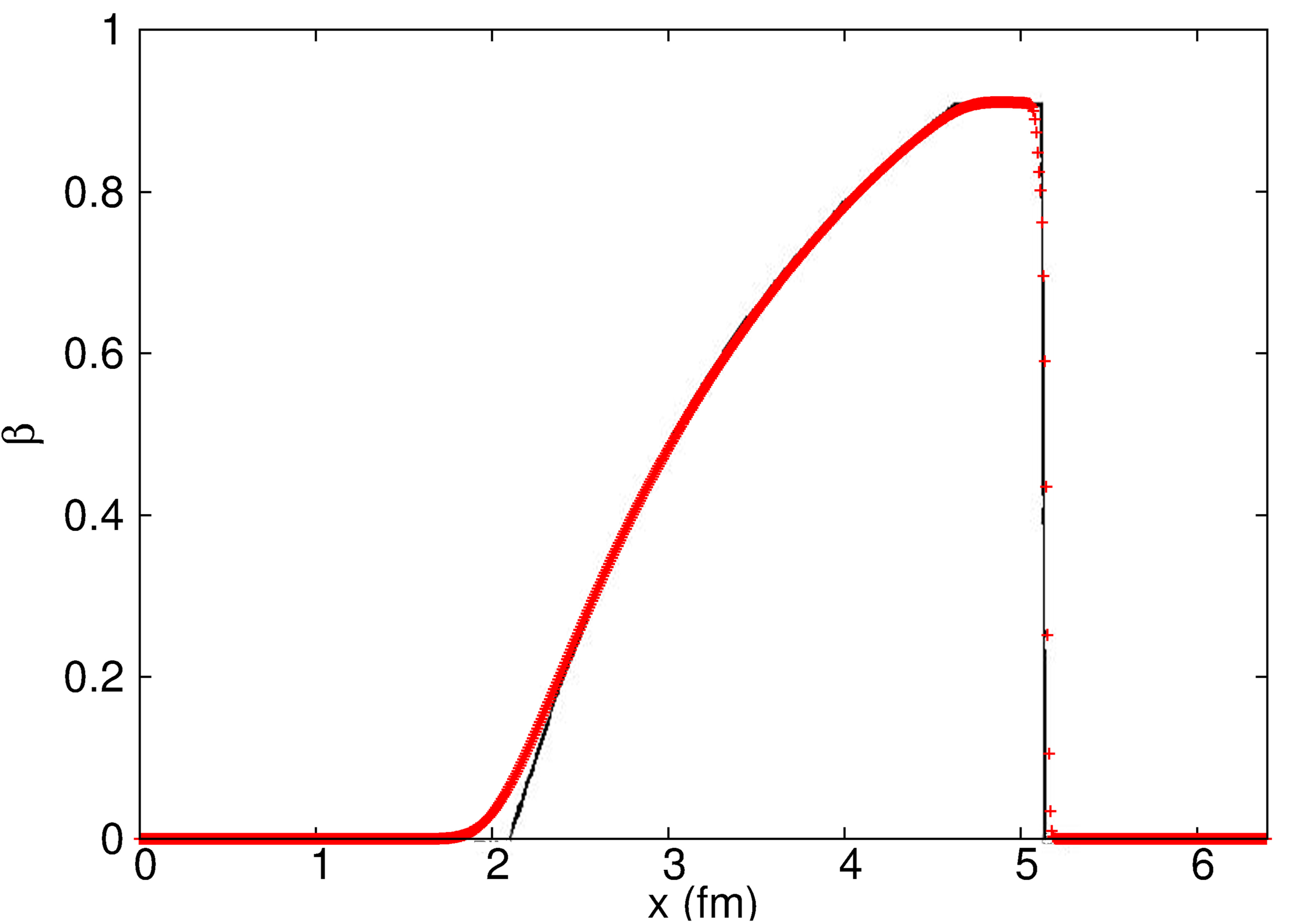}
\end{center}
\caption{Comparison between the extended model and analytical results
for the pressure (top) and velocity (bottom) profiles in the highly relativistic regime. }
\label{meanalytical}
\end{figure}

\begin{figure}[h]
\begin{center}
\includegraphics[trim=10mm 20mm 10mm 10mm, clip, width=0.9\columnwidth, height=0.6\columnwidth ]{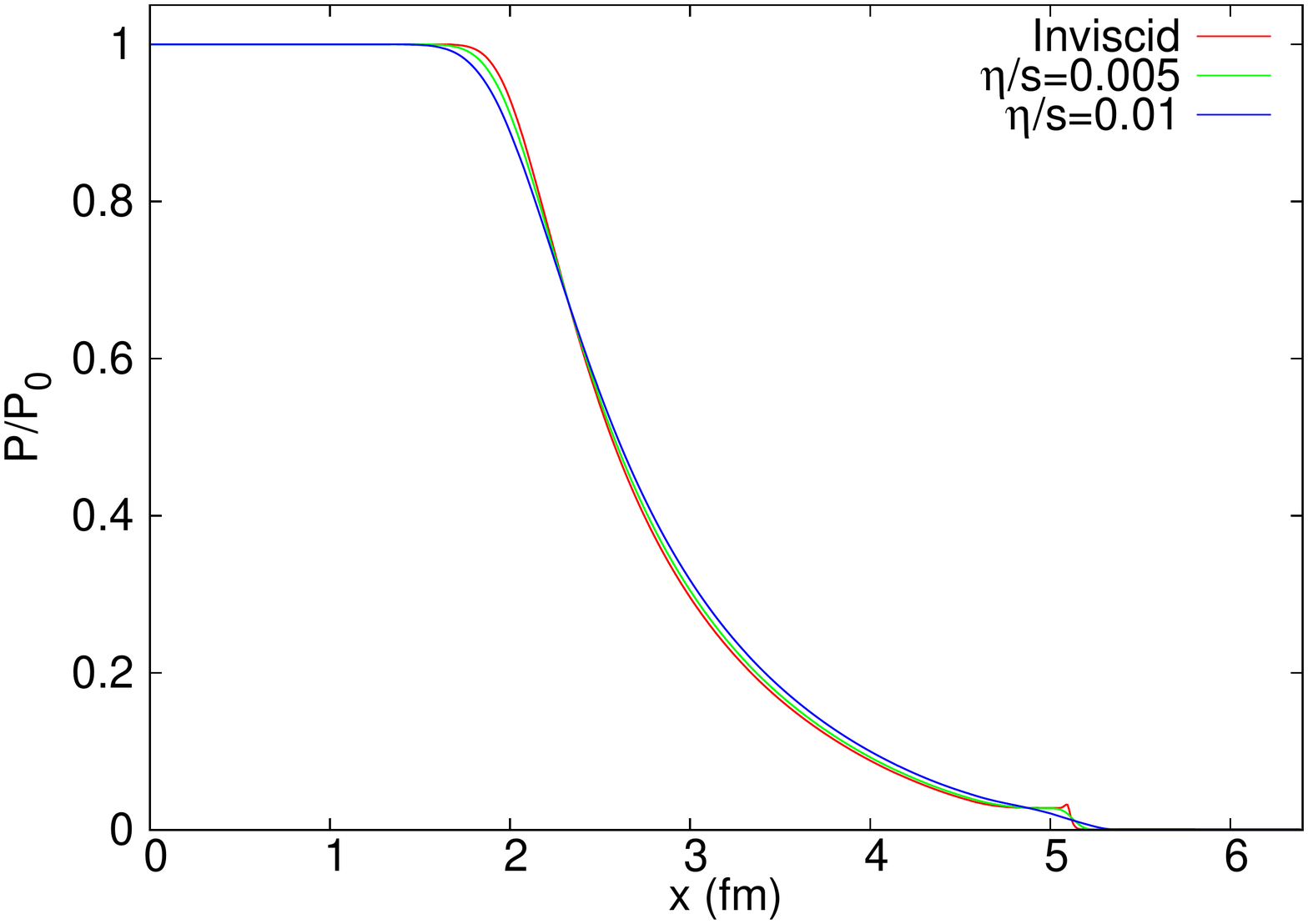}
\includegraphics[trim=10mm 20mm 10mm 10mm, clip, width=0.9\columnwidth, height=0.6\columnwidth ]{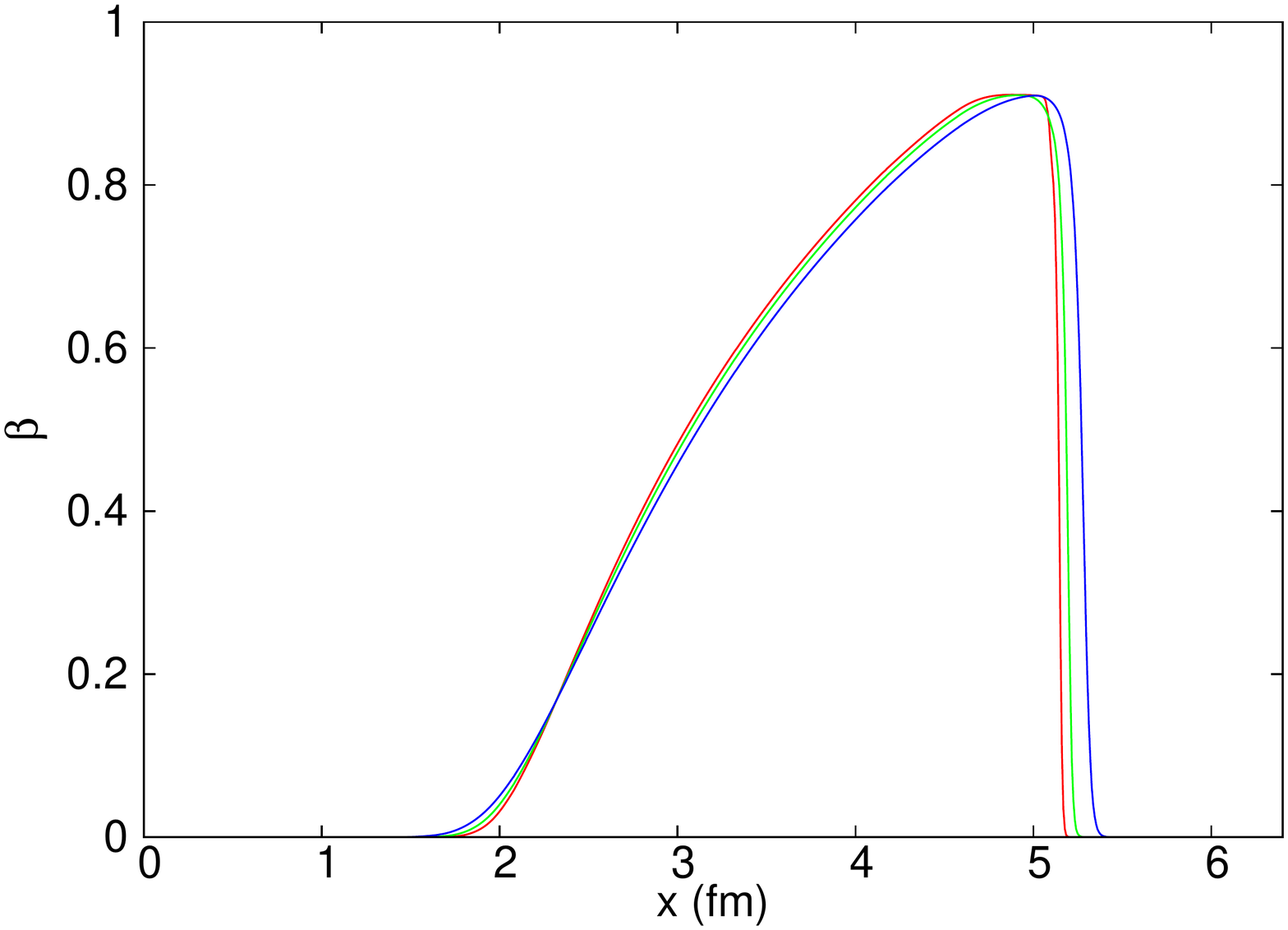}
\end{center}
\caption{Comparison of the results for the inviscid case and different
 $\eta/s$, for the pressure (top) and velocity (bottom) profiles in the
 highly relativistic regime. }
\label{highviscosity}
\end{figure}

\begin{figure}[h]
\begin{center}
\includegraphics[trim=10mm 20mm 10mm 10mm, clip, width=0.9\columnwidth, height=0.6\columnwidth ]{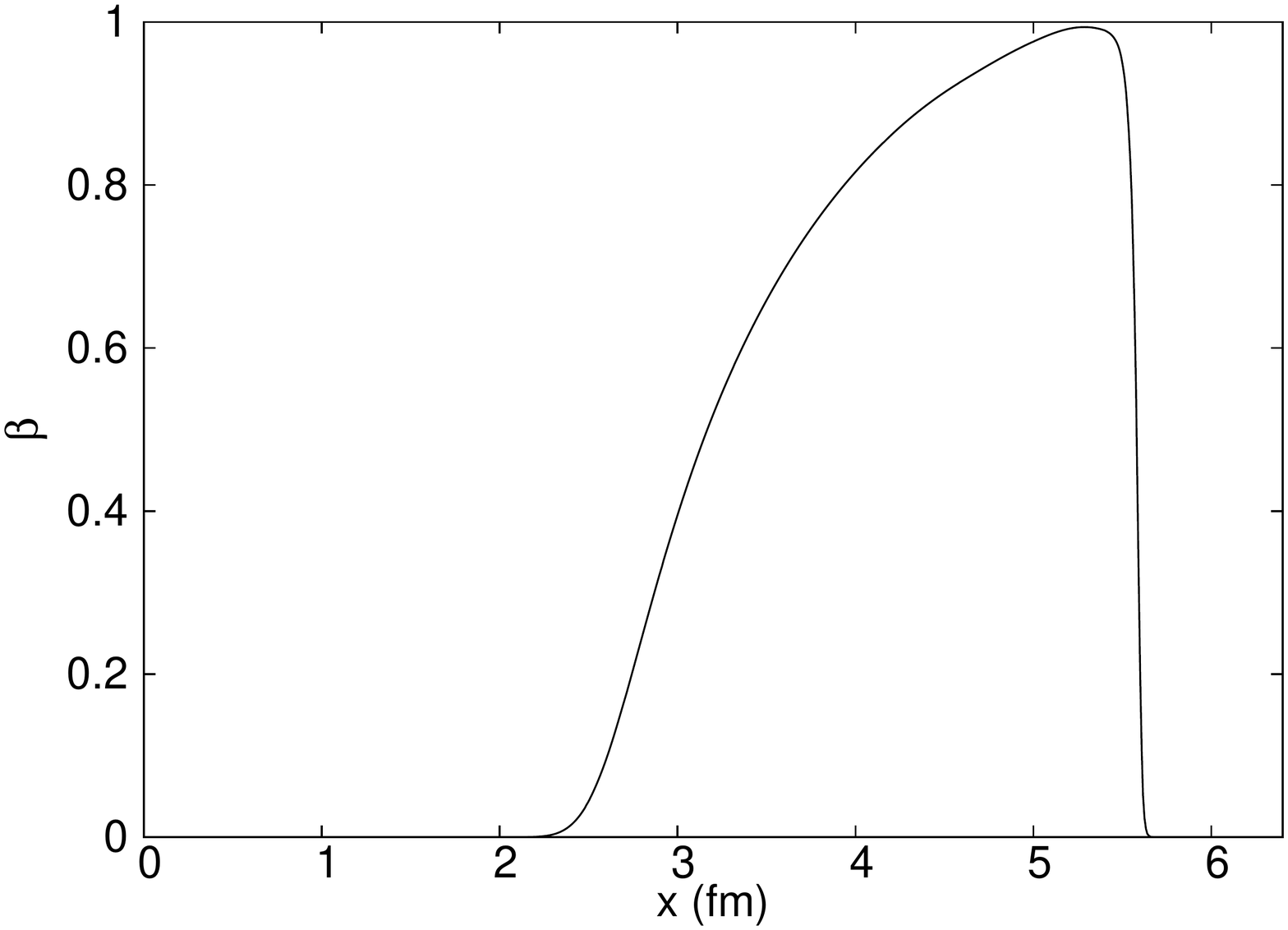}
\includegraphics[trim=10mm 20mm 10mm 10mm, clip, width=0.9\columnwidth, height=0.6\columnwidth ]{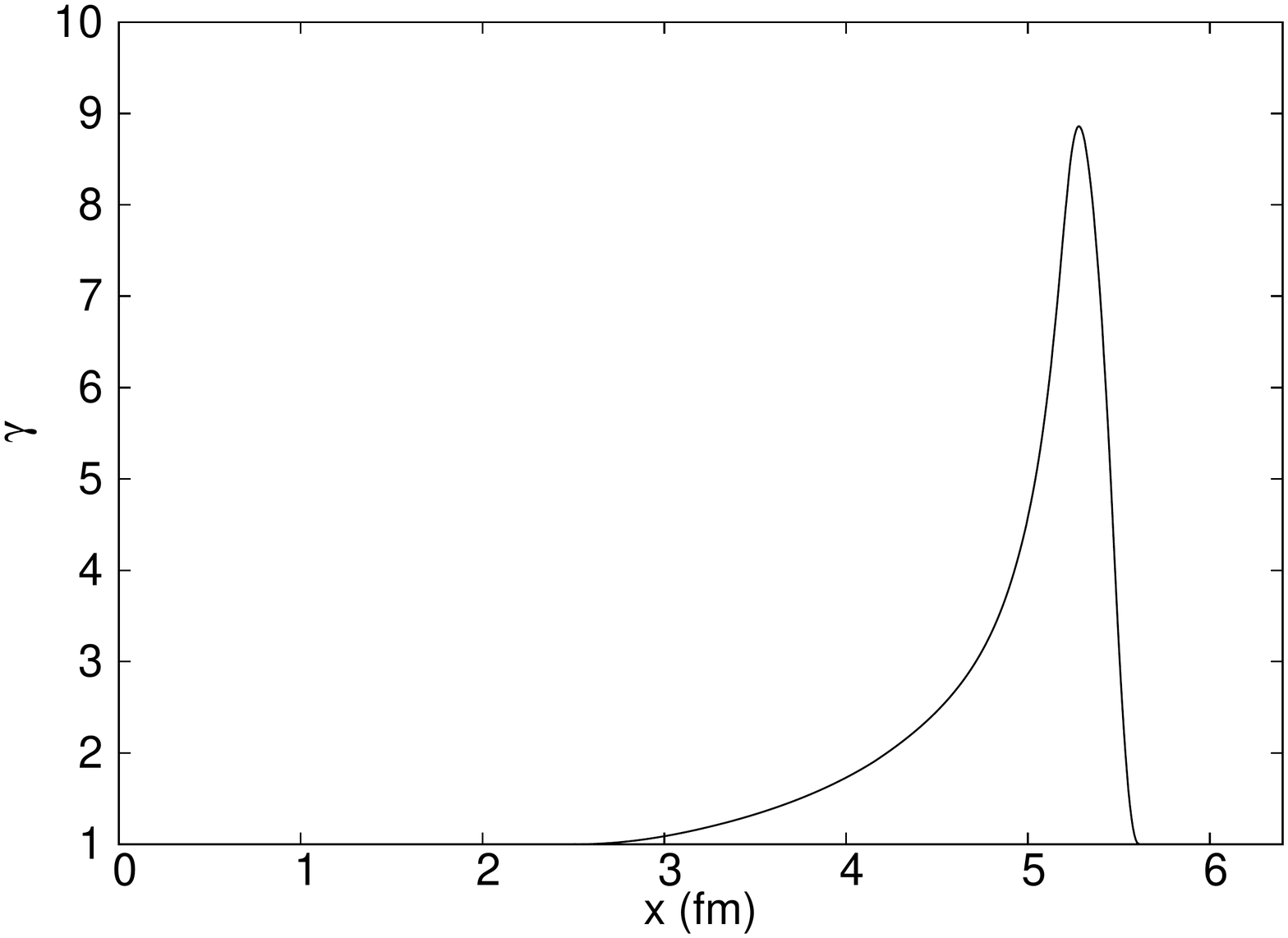}
\end{center}
\caption{Results of the simulation for the velocity profile (top) and
  Lorentz's factor (bottom) in the ultra-high relativistic regime. }
\label{highvelocity}
\end{figure}

\begin{figure}[h]
\begin{center}
\includegraphics[trim=20mm 90mm 10mm 85mm, clip, scale=0.45]{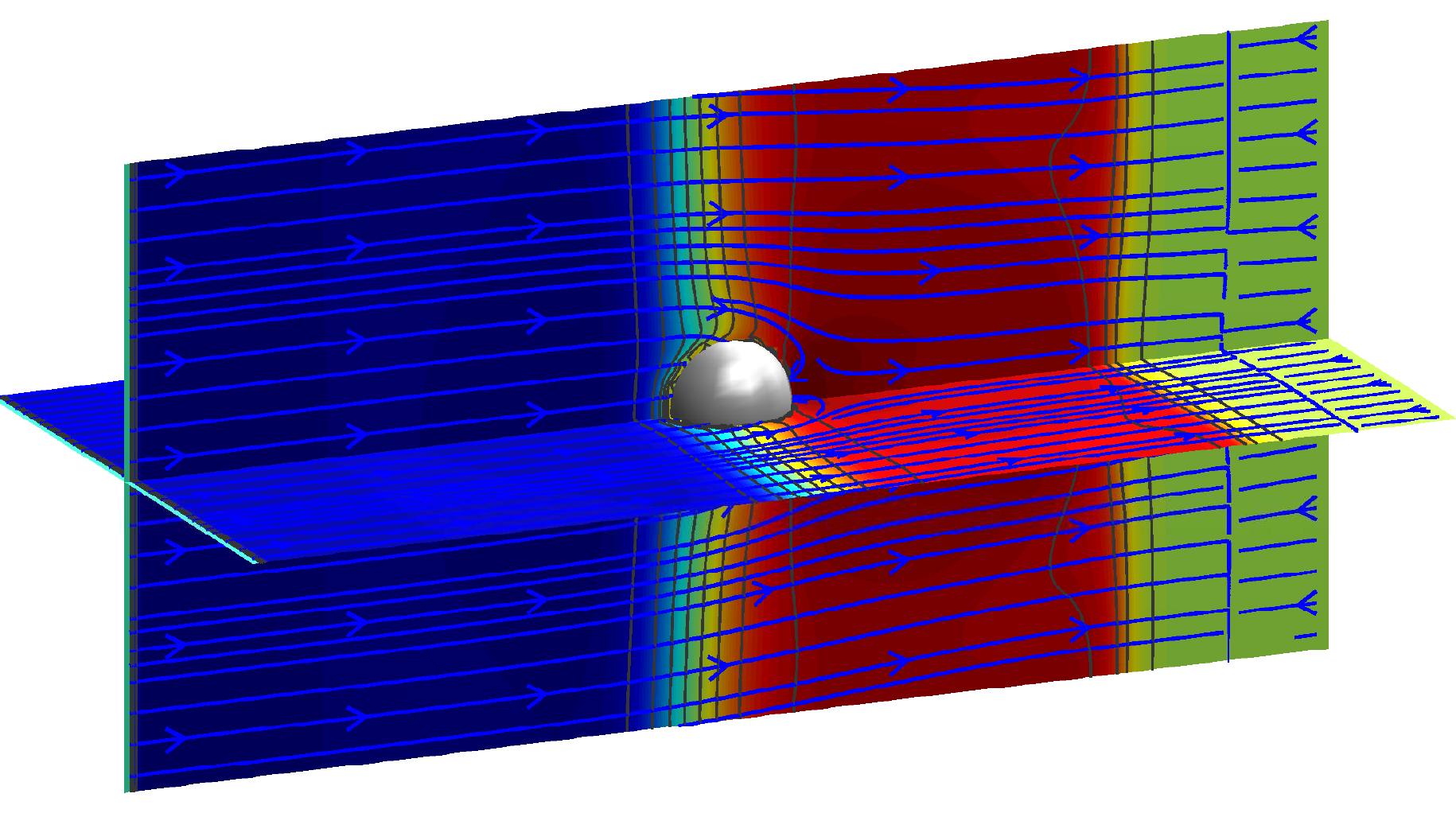}
\includegraphics[trim=25mm 90mm 10mm 85mm, clip, scale=0.48]{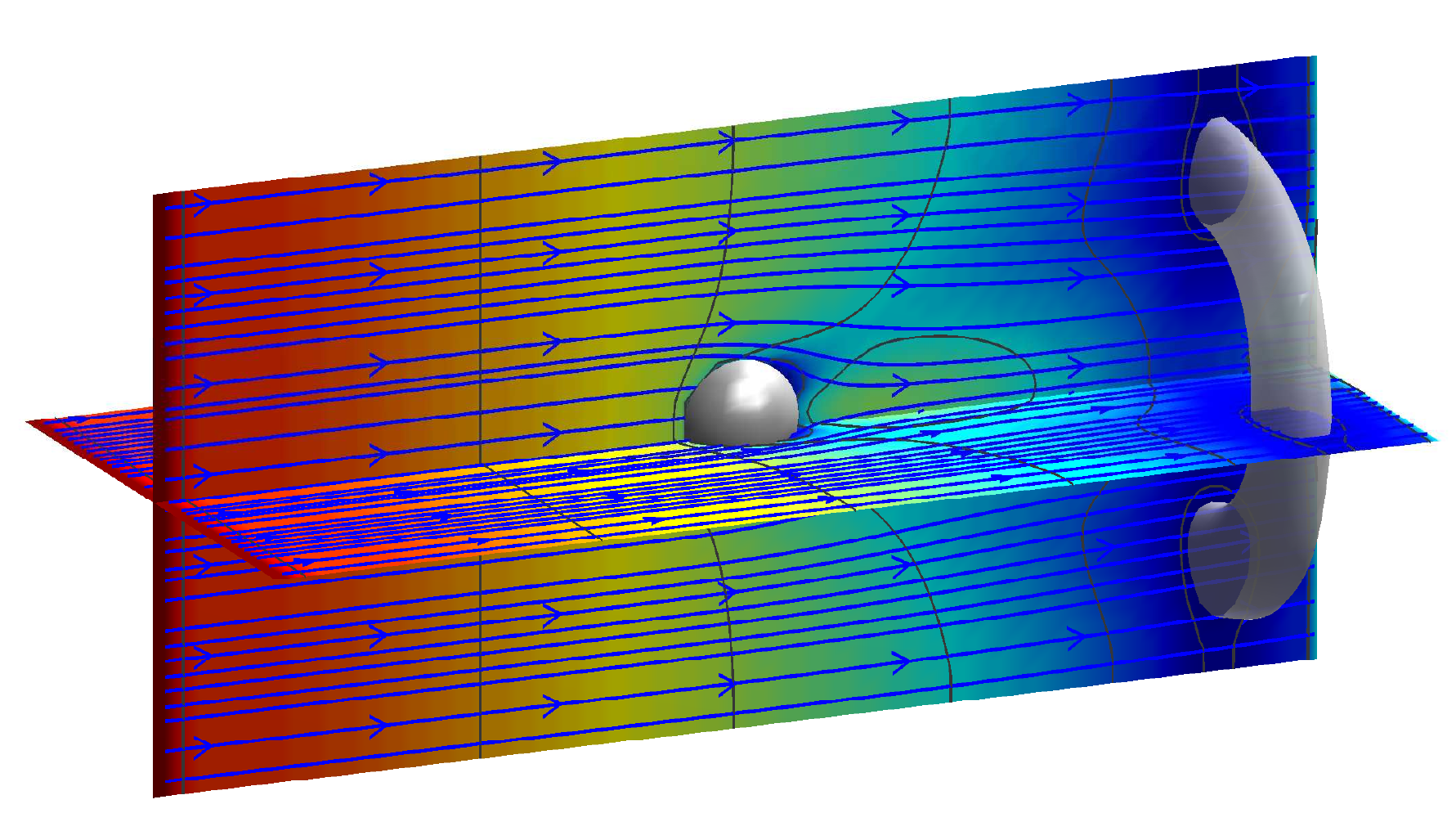}
\end{center}
\caption{Snapshots of the three-dimensional simulation of a relativistic
  shock wave colliding with a massive interstellar cloud. Here, the density field is plotted in logarithmic scale in the weakly
  relativistic regime (top) and highly relativistic regime (bottom) at
  time $t=1000$. The iso-surface in the second figure illustrates a
  region of low density ($\log(n/n_0) \sim -2.5$).}
\label{super3D}
\end{figure}

\begin{figure}[h]
\begin{center}
\includegraphics[trim=0mm 0mm 0mm 0mm, clip, width=0.9\columnwidth, height=0.55\columnwidth ]{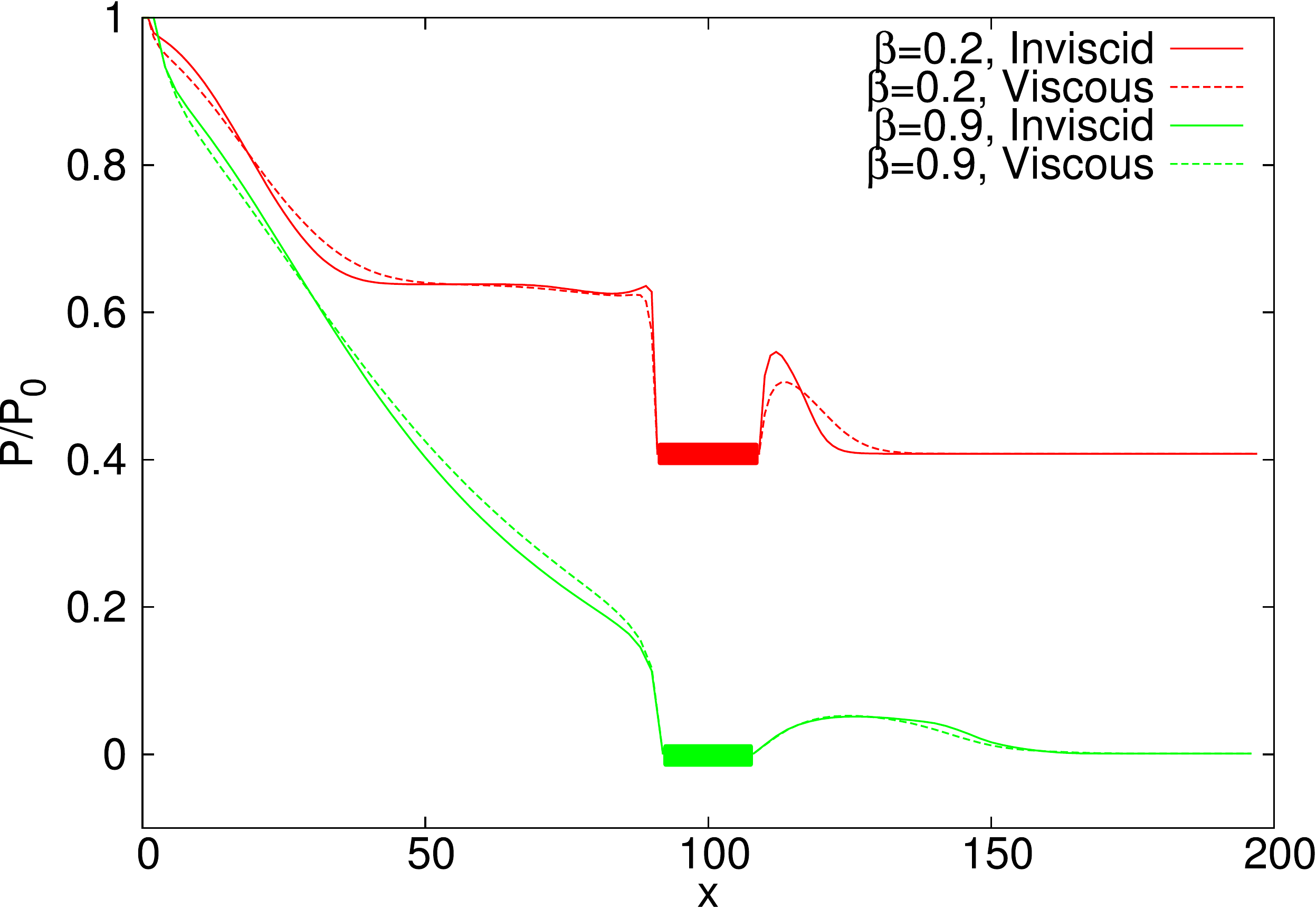}
\includegraphics[trim=0mm 0mm 0mm 0mm, clip, width=0.9\columnwidth, height=0.55\columnwidth ]{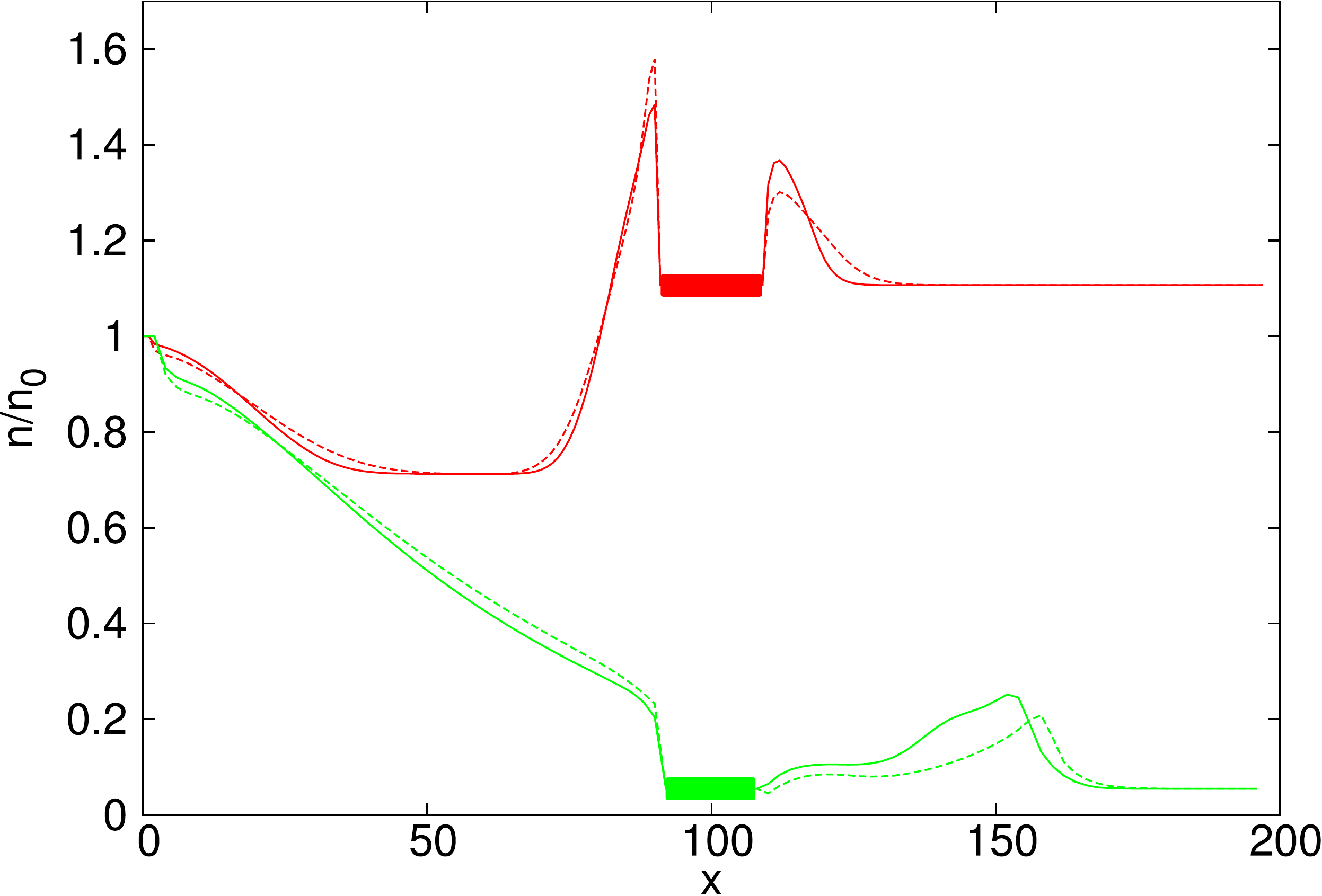}
\includegraphics[trim=0mm 0mm 0mm 0mm, clip, width=0.9\columnwidth, height=0.55\columnwidth ]{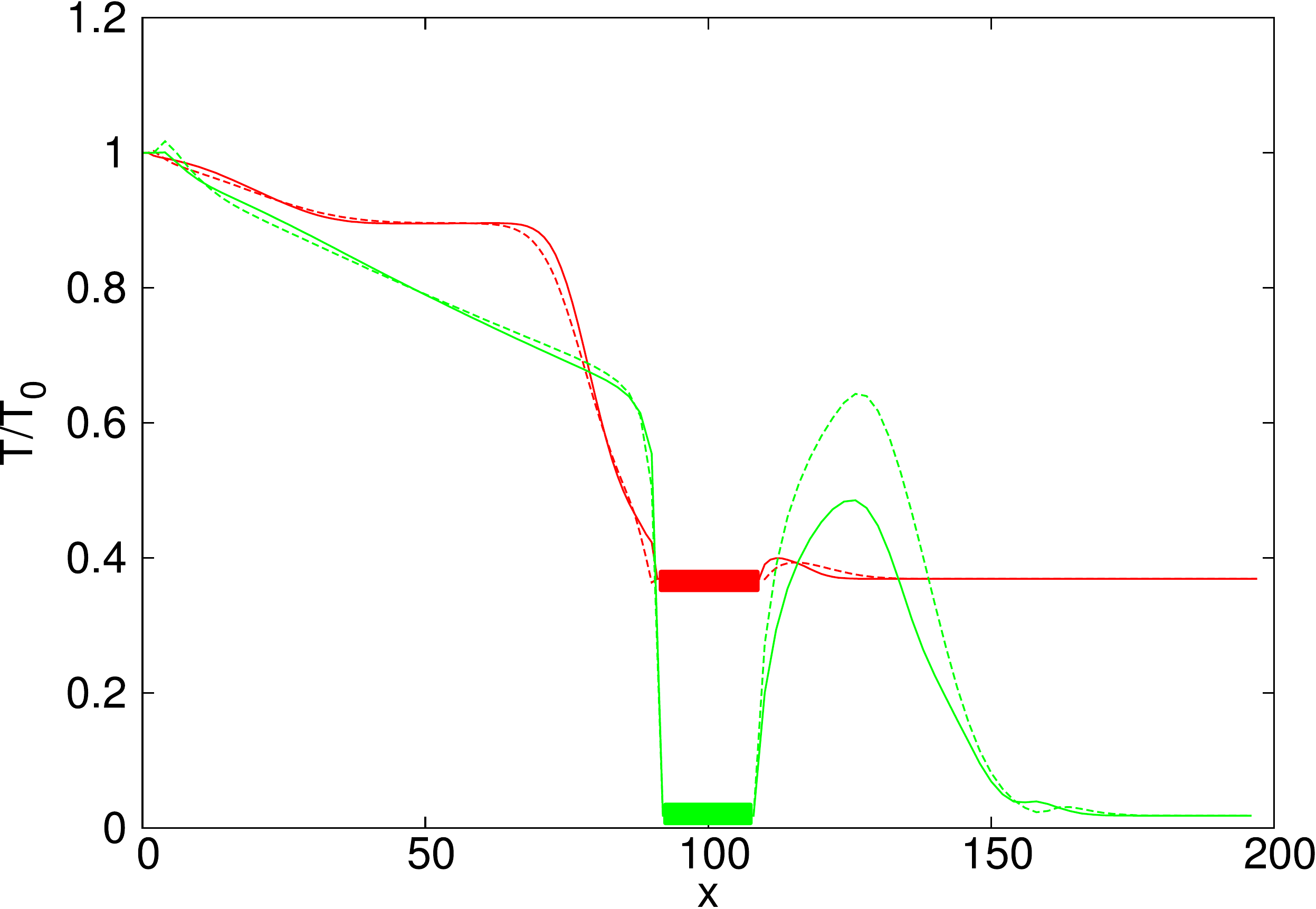}
\end{center}
\caption{Pressure (top), density (middle) and temperature (bottom) profiles in the weakly and highly relativistic regimes, for the inviscid and viscous case. The results are shown along the $x$ axis at $y=z=50$ at $t=600$ time steps. The thicker lines denote the regions where the interstellar cloud is located. }
\label{super2D}
\end{figure}

The results of our inviscid simulation are compared to the analytical
results in Fig.~\ref{meanalytical}. To drive the shock at
velocity $\beta \sim 0.9$ (highly relativistic regime), the applied
pressure is set to $P_0=5.43 GeV /fm^3$ and $P_1=5.43 MeV/fm^3$, which
corresponds to $ 2.495\times10^{-7}$ and $2.495\times10^{-10}$ in
numerical units, respectively. The results are shown at $t=2.0 fm/s$
and very good agreement is found. The small discrepancies between our
simulation and the analytical curve are related to the fact that a
small value of bulk viscosity inevitably remains in our simulation, due to the
$\lambda_i$ coefficients, which are needed to increase the stability
of the model. This issue makes a subject for future investigations.
The CPU time required for the
simulation shown in Fig.~\ref{meanalytical} is 593 ms.

The fact that we can model properly the Euler equation at
very high velocities opens the possibility of using our model in
astrophysical applications, where velocities are usually high and
viscous effects are negligible. 

Fig.~\ref{highviscosity} shows the same result at $\beta \sim 0.9$ for
different viscosities, compared to the inviscid case at $t=2.0
fm/s$. The same conditions as mentioned above are considered
again. The effects of increasing $\eta/s$ at this velocity are
similar to the case of lower velocity.

In order to demonstrate the ability of the model to simulate
ultra-high velocities, we consider the case of the initial pressures
$P_0=5.43 GeV /fm^3$ and $P_1=0.0543 MeV/fm^3$ which corresponds to
$2.495\times10^{-7}$ and $ 2.495\times10^{-12}$ in numerical units,
respectively. We set $\delta_t/\delta_x=0.25$, $\alpha=0.2$, and
$\eta/s=0.01$. The results for the velocity profile and local
Lorentz's factor at $t=2.0fm/s$ are presented in
Fig.~\ref{highvelocity}, which shows that for this case $\beta \sim
0.99$ and $\gamma(u) \sim 9$. This indicates that our model is
numerically stable for simulating relativistic fluids with ultra-high
velocities. The required CPU time for this simulation is $781$ ms.

\subsection*{Astrophysical application}

As an astrophysical application, we simulate a relativistic shock
wave, generated by, say, a supernova explosion, colliding with
a massive interstellar cloud, e.g. molecular gas
\cite{scienceSuper}. The ejecta from the explosion of such supernovae
are known to sweep the interstellar material up to relativistic
velocities along the way (relativistic outflows)
\cite{natureSuper}. We perform a three-dimensional simulation of a
shock wave passing through a cold spherical cloud in a lattice of
$200\times100\times100$ cells. As mentioned earlier, in order to solve
the equation of conservation of particle density, an extra
distribution function is used, Eq.\eqref{Hupp}.  As initial
condition, the region is divided in two zones by the plane $x=65$; at
the left hand side ($x \leq 65$), the density is set to $n_1=0.6
cm^{-3}$ and the temperature to $T_1=10^4 K$. The massive cloud is
modeled as a solid sphere with radius of $10$ cells and centered at the
location $(100,50,50)$, where we neglect the drag force acting on the
cloud due to the flow (the sphere will remain at the same position
during the whole simulation). Open boundary conditions are applied to
all the external boundaries, except the left one, where an inlet
boundary condition is applied by fixing the distribution function with
the equilibrium distribution calculated at the initial condition,
i.e. $n_0$ and $T_0$. On the surface of the cloud, the cells evolve to the equilibrium distribution function evaluated at the
constant values of $n=n_1$, $T=T_1$ and $\boldsymbol{u}=0$. 
It should be mentioned that, at each cell, the pressure can be calculated using
the relation $P=nT$.

By changing the initial condition at the right hand side of the
dividing plane ($x>65$), we are able to tune the velocity of the shock
wave. Two different velocities are considered here. For the first
case, we consider a shock wave at weakly relativistic regime
($\beta=0.2$), setting $T_0=2.71 T_1$ and $n_0=0.9 n_1$. For the
second case, highly relativistic regime ($\beta=0.9$), theses values
are $T_0=54.77 T_1$ and $n_0=10.95 n_1$. Fig.~\ref{super3D} shows the
results of the 3D simulation of the shock wave, after colliding with
the massive interstellar cloud for the density field for both
cases. The simulations are performed in the inviscid case, where
$\alpha=0.2$ and $\delta_t/\delta_x=0.15$. The density field
is plotted in logarithmic scale at $t=1000$ time step, where red and
blue denote high and low values, respectively, and streamlines
represent the velocity field. In this figure, we observe an increase
in the density downstream of the collision, likely due to the
fact that, during the propagation, the shock wave collects
interstellar material and pushes it against the cloud (sweeping
effect). However, at higher speeds, this increase on the density
becomes less pronounced, and the passing of the shock wave generates a
ring-shape region of low density downstream (see isosurface in
Fig.~\ref{super3D}). 

In order to study the viscous effects, the same simulations are
performed by introducing the dissipation and taking
$\tau=1$. Fig.~\ref{super2D} shows the pressure, density and
temperature profiles at weakly and highly relativistic regimes for
both, inviscid and viscous cases. The results are shown along the $x $
axis at $y=z=50$ and at $t=600$ time steps. Note that
  since $P_0$, $n_0$, and $T_0$ take very different values for the
  weakly and highly relativistic regimes, in order to make a clear
  comparison of the results, we have normalized $P$, $n$, and $T$ with
  $P_0$, $n_0$, and $T_0$, respectively. As it can be appreciated,
the viscous effects become relatively more important downstream of the
cloud. Furthermore, it causes a pressure drop and decreases the
density, while inducing a corresponding increment of temperature in
the gas. Note, that the effects on the temperature are more
significant at highly relativistic regime than the pressure drop,
while an opposite behavior is found at weakly relativistic regime.

\section {Conclusions}
\label {Conclusions}

In this paper, we have introduced a relativistic lattice Boltzmann
model that is able to handle relativistic fluid dynamics at very high velocities.
For this purpose, we have first expanded the Maxwell J\"uttner
distribution in orthogonal polynomials, by assuming as weight function
the equilibrium distribution at the local rest frame. A discretization
procedure has been applied in order to adjust the expansion to the
D3Q19 cell configuration, which, in order to avoid a multi-time
evolution of the Boltzmann equation, leads to the problem of
recovering only the conservation of the momentum-energy
tensor. However, in the ultra-relativistic regime ($\epsilon = 3p$) the
entire dynamics of the system is governed by this equation and the
first order moment is not required. To extend the model to high
velocities ($\beta \sim 1$), we use a flux limiter scheme and
introduce a bulk viscosity term into the Boltzmann equation, to
increase the numerical stability in presence of discontinuities.

In order to validate our model, we have compared the numerical results
for shock waves in viscous quark-gluon plasmas with the results of
other existing models, and found very good agreement. 
In addition, to the best of our knowledge,  we have for the first
time successfully simulated shock waves in relativistic viscous flow for $\beta
>0.6$. We have also suggested a way to simulate near-inviscid flows (Euler
equation) using the extended model by modifying the collision
step. For this case, we have compared the results with the analytical
solution, finding again very satisafctory agreement. 
This offers a promising strategy  to study astrophysical flows 
at very high speeds and negligible viscous effects. 
Additionally, we have shown that our model is capable of simulating 
the Riemann problem at ultra-high relativistic flows ($\gamma \sim 10$). 
Finally, we have studied the collision of a shock wave colliding with a massive
interstellar cloud in weakly and highly relativistic regimes, for both
inviscid and viscous cases.

Summarizing, we have proposed a model which is simple,
numerically efficient, and capable of simulating highly relativistic
flows. It can also be used to simulate near-inviscid flows. Moreover, like
other lattice Boltzmann methods, our model is highly adaptable to
parallel computing and can be used to simulate complex
geometries. These features make this model appealing for 
prospective applications in relativistic astrophysics and high energy physics. 
Extensions of this model to include higher order lattices so as to
recover more moments of the equilibrium distribution, make a very
interesting subject for future research.

\begin{acknowledgments}
  We acknowledge financial support from the European Research Council
  (ERC) Advanced Grant 319968-FlowCCS and financial support of the Eidgenössische Technische
  Hochschule Z\"urich (ETHZ) under Grant No. 0611-1.
\end{acknowledgments}
\bibliography{references}

\begin{thebibliography}{38}%
\makeatletter
\providecommand \@ifxundefined [1]{%
 \@ifx{#1\undefined}
}%
\providecommand \@ifnum [1]{%
 \ifnum #1\expandafter \@firstoftwo
 \else \expandafter \@secondoftwo
 \fi
}%
\providecommand \@ifx [1]{%
 \ifx #1\expandafter \@firstoftwo
 \else \expandafter \@secondoftwo
 \fi
}%
\providecommand \natexlab [1]{#1}%
\providecommand \enquote  [1]{``#1''}%
\providecommand \bibnamefont  [1]{#1}%
\providecommand \bibfnamefont [1]{#1}%
\providecommand \citenamefont [1]{#1}%
\providecommand \href@noop [0]{\@secondoftwo}%
\providecommand \href [0]{\begingroup \@sanitize@url \@href}%
\providecommand \@href[1]{\@@startlink{#1}\@@href}%
\providecommand \@@href[1]{\endgroup#1\@@endlink}%
\providecommand \@sanitize@url [0]{\catcode `\\12\catcode `\$12\catcode
  `\&12\catcode `\#12\catcode `\^12\catcode `\_12\catcode `\%12\relax}%
\providecommand \@@startlink[1]{}%
\providecommand \@@endlink[0]{}%
\providecommand \url  [0]{\begingroup\@sanitize@url \@url }%
\providecommand \@url [1]{\endgroup\@href {#1}{\urlprefix }}%
\providecommand \urlprefix  [0]{URL }%
\providecommand \Eprint [0]{\href }%
\providecommand \doibase [0]{http://dx.doi.org/}%
\providecommand \selectlanguage [0]{\@gobble}%
\providecommand \bibinfo  [0]{\@secondoftwo}%
\providecommand \bibfield  [0]{\@secondoftwo}%
\providecommand \translation [1]{[#1]}%
\providecommand \BibitemOpen [0]{}%
\providecommand \bibitemStop [0]{}%
\providecommand \bibitemNoStop [0]{.\EOS\space}%
\providecommand \EOS [0]{\spacefactor3000\relax}%
\providecommand \BibitemShut  [1]{\csname bibitem#1\endcsname}%
\let\auto@bib@innerbib\@empty
\bibitem [{\citenamefont {Blandford}(1974)}]{relativisticjet}%
  \BibitemOpen
  \bibfield  {author} {\bibinfo {author} {\bibfnamefont {M.~J.}\ \bibnamefont
  {Blandford}, \bibfnamefont {R.~D.;~Rees}},\ }\href@noop {} {\bibfield
  {journal} {\bibinfo  {journal} {MNRAS}\ }\textbf {\bibinfo {volume} {169}},\
  \bibinfo {pages} {395} (\bibinfo {year} {1974})}\BibitemShut {NoStop}%
\bibitem [{\citenamefont {Nishihara}\ \emph {et~al.}(2010)\citenamefont
  {Nishihara}, \citenamefont {Wouchuk}, \citenamefont {Matsuoka}, \citenamefont
  {Ishizaki},\ and\ \citenamefont {Zhakhovsky}}]{nishihara2010richtmyer}%
  \BibitemOpen
  \bibfield  {author} {\bibinfo {author} {\bibfnamefont {K.}~\bibnamefont
  {Nishihara}}, \bibinfo {author} {\bibfnamefont {J.}~\bibnamefont {Wouchuk}},
  \bibinfo {author} {\bibfnamefont {C.}~\bibnamefont {Matsuoka}}, \bibinfo
  {author} {\bibfnamefont {R.}~\bibnamefont {Ishizaki}}, \ and\ \bibinfo
  {author} {\bibfnamefont {V.}~\bibnamefont {Zhakhovsky}},\ }\href@noop {}
  {\bibfield  {journal} {\bibinfo  {journal} {Philosophical Transactions of the
  Royal Society A: Mathematical, Physical and Engineering Sciences}\ }\textbf
  {\bibinfo {volume} {368}},\ \bibinfo {pages} {1769} (\bibinfo {year}
  {2010})}\BibitemShut {NoStop}%
\bibitem [{\citenamefont {Shuryak}(2004)}]{shuryak2004does}%
  \BibitemOpen
  \bibfield  {author} {\bibinfo {author} {\bibfnamefont {E.}~\bibnamefont
  {Shuryak}},\ }\href@noop {} {\bibfield  {journal} {\bibinfo  {journal}
  {Progress in Particle and Nuclear Physics}\ }\textbf {\bibinfo {volume}
  {53}},\ \bibinfo {pages} {273} (\bibinfo {year} {2004})}\BibitemShut
  {NoStop}%
\bibitem [{\citenamefont {Dubal}(1991)}]{dubal1991numerical}%
  \BibitemOpen
  \bibfield  {author} {\bibinfo {author} {\bibfnamefont {M.}~\bibnamefont
  {Dubal}},\ }\href@noop {} {\bibfield  {journal} {\bibinfo  {journal}
  {Computer Physics Communications}\ }\textbf {\bibinfo {volume} {64}},\
  \bibinfo {pages} {221} (\bibinfo {year} {1991})}\BibitemShut {NoStop}%
\bibitem [{\citenamefont {Hernquist}\ and\ \citenamefont
  {Katz}(1989)}]{TREESPH}%
  \BibitemOpen
  \bibfield  {author} {\bibinfo {author} {\bibfnamefont {L.}~\bibnamefont
  {Hernquist}}\ and\ \bibinfo {author} {\bibfnamefont {N.}~\bibnamefont
  {Katz}},\ }\href@noop {} {\bibfield  {journal} {\bibinfo  {journal}
  {Astrophysical Journal Supplement Series}\ }\textbf {\bibinfo {volume}
  {70}},\ \bibinfo {pages} {419} (\bibinfo {year} {1989})}\BibitemShut
  {NoStop}%
\bibitem [{\citenamefont {Siegler}\ and\ \citenamefont
  {Riffert}(2008)}]{siegler2008smoothed}%
  \BibitemOpen
  \bibfield  {author} {\bibinfo {author} {\bibfnamefont {S.}~\bibnamefont
  {Siegler}}\ and\ \bibinfo {author} {\bibfnamefont {H.}~\bibnamefont
  {Riffert}},\ }\href@noop {} {\bibfield  {journal} {\bibinfo  {journal} {The
  Astrophysical Journal}\ }\textbf {\bibinfo {volume} {531}},\ \bibinfo {pages}
  {1053} (\bibinfo {year} {2008})}\BibitemShut {NoStop}%
\bibitem [{\citenamefont {Wen}\ \emph {et~al.}(1997)\citenamefont {Wen},
  \citenamefont {Panaitescu},\ and\ \citenamefont {Laguna}}]{wen1997shock}%
  \BibitemOpen
  \bibfield  {author} {\bibinfo {author} {\bibfnamefont {L.}~\bibnamefont
  {Wen}}, \bibinfo {author} {\bibfnamefont {A.}~\bibnamefont {Panaitescu}}, \
  and\ \bibinfo {author} {\bibfnamefont {P.}~\bibnamefont {Laguna}},\
  }\href@noop {} {\bibfield  {journal} {\bibinfo  {journal} {The Astrophysical
  Journal}\ }\textbf {\bibinfo {volume} {486}},\ \bibinfo {pages} {919}
  (\bibinfo {year} {1997})}\BibitemShut {NoStop}%
\bibitem [{\citenamefont {Eulderink}\ and\ \citenamefont
  {Mellema}(1995)}]{Generalrelativistichydrodynamics}%
  \BibitemOpen
  \bibfield  {author} {\bibinfo {author} {\bibfnamefont {F.}~\bibnamefont
  {Eulderink}}\ and\ \bibinfo {author} {\bibfnamefont {G.}~\bibnamefont
  {Mellema}},\ }\href@noop {} {\bibfield  {journal} {\bibinfo  {journal}
  {Astronomy and Astrophysics Supplement}\ }\textbf {\bibinfo {volume} {110}},\
  \bibinfo {pages} {110} (\bibinfo {year} {1995})}\BibitemShut {NoStop}%
\bibitem [{\citenamefont {Yang}\ \emph {et~al.}(1997)\citenamefont {Yang},
  \citenamefont {Chen}, \citenamefont {Tsai},\ and\ \citenamefont
  {Chang}}]{yang1997kinetic}%
  \BibitemOpen
  \bibfield  {author} {\bibinfo {author} {\bibfnamefont {J.}~\bibnamefont
  {Yang}}, \bibinfo {author} {\bibfnamefont {M.}~\bibnamefont {Chen}}, \bibinfo
  {author} {\bibfnamefont {I.}~\bibnamefont {Tsai}}, \ and\ \bibinfo {author}
  {\bibfnamefont {J.}~\bibnamefont {Chang}},\ }\href@noop {} {\bibfield
  {journal} {\bibinfo  {journal} {Journal of Computational Physics}\ }\textbf
  {\bibinfo {volume} {136}},\ \bibinfo {pages} {19} (\bibinfo {year}
  {1997})}\BibitemShut {NoStop}%
\bibitem [{\citenamefont {Benzi}\ \emph {et~al.}(1992)\citenamefont {Benzi},
  \citenamefont {Succi},\ and\ \citenamefont {Vergassola}}]{Benzi1992145}%
  \BibitemOpen
  \bibfield  {author} {\bibinfo {author} {\bibfnamefont {R.}~\bibnamefont
  {Benzi}}, \bibinfo {author} {\bibfnamefont {S.}~\bibnamefont {Succi}}, \ and\
  \bibinfo {author} {\bibfnamefont {M.}~\bibnamefont {Vergassola}},\ }\href
  {\doibase 10.1016/0370-1573(92)90090-M} {\bibfield  {journal} {\bibinfo
  {journal} {Physics Reports}\ }\textbf {\bibinfo {volume} {222}},\ \bibinfo
  {pages} {145 } (\bibinfo {year} {1992})}\BibitemShut {NoStop}%
\bibitem [{\citenamefont {Chen}\ \emph {et~al.}(1992)\citenamefont {Chen},
  \citenamefont {Chen},\ and\ \citenamefont {Matthaeus}}]{chen1992recovery}%
  \BibitemOpen
  \bibfield  {author} {\bibinfo {author} {\bibfnamefont {H.}~\bibnamefont
  {Chen}}, \bibinfo {author} {\bibfnamefont {S.}~\bibnamefont {Chen}}, \ and\
  \bibinfo {author} {\bibfnamefont {W.}~\bibnamefont {Matthaeus}},\ }\href@noop
  {} {\bibfield  {journal} {\bibinfo  {journal} {Physical Review A}\ }\textbf
  {\bibinfo {volume} {45}},\ \bibinfo {pages} {5339} (\bibinfo {year}
  {1992})}\BibitemShut {NoStop}%
\bibitem [{\citenamefont {Succi}(2001)}]{succi2001lattice}%
  \BibitemOpen
  \bibfield  {author} {\bibinfo {author} {\bibfnamefont {S.}~\bibnamefont
  {Succi}},\ }\href@noop {} {\emph {\bibinfo {title} {The lattice Boltzmann
  equation for Fluid Dynamics and Beyond}}}\ (\bibinfo  {publisher} {Oxford
  University Press},\ \bibinfo {address} {New York},\ \bibinfo {year}
  {2001})\BibitemShut {NoStop}%
\bibitem [{\citenamefont {Succi}\ \emph {et~al.}(1997)\citenamefont {Succi},
  \citenamefont {Wang},\ and\ \citenamefont
  {Qian}}]{doi:10.1142/S0129183197000862}%
  \BibitemOpen
  \bibfield  {author} {\bibinfo {author} {\bibfnamefont {S.}~\bibnamefont
  {Succi}}, \bibinfo {author} {\bibfnamefont {J.}~\bibnamefont {Wang}}, \ and\
  \bibinfo {author} {\bibfnamefont {Y.-H.}\ \bibnamefont {Qian}},\ }\href
  {\doibase 10.1142/S0129183197000862} {\bibfield  {journal} {\bibinfo
  {journal} {International Journal of Modern Physics C}\ }\textbf {\bibinfo
  {volume} {08}},\ \bibinfo {pages} {999} (\bibinfo {year} {1997})}\BibitemShut
  {NoStop}%
\bibitem [{\citenamefont {McNamara}\ and\ \citenamefont
  {Zanetti}(1988)}]{PhysRevLett.61.2332}%
  \BibitemOpen
  \bibfield  {author} {\bibinfo {author} {\bibfnamefont {G.~R.}\ \bibnamefont
  {McNamara}}\ and\ \bibinfo {author} {\bibfnamefont {G.}~\bibnamefont
  {Zanetti}},\ }\href {\doibase 10.1103/PhysRevLett.61.2332} {\bibfield
  {journal} {\bibinfo  {journal} {Phys. Rev. Lett.}\ }\textbf {\bibinfo
  {volume} {61}},\ \bibinfo {pages} {2332} (\bibinfo {year}
  {1988})}\BibitemShut {NoStop}%
\bibitem [{\citenamefont {Higuera}\ and\ \citenamefont
  {Jimenez}(2007)}]{higuera2007boltzmann}%
  \BibitemOpen
  \bibfield  {author} {\bibinfo {author} {\bibfnamefont {F.}~\bibnamefont
  {Higuera}}\ and\ \bibinfo {author} {\bibfnamefont {J.}~\bibnamefont
  {Jimenez}},\ }\href@noop {} {\bibfield  {journal} {\bibinfo  {journal} {EPL
  (Europhysics Letters)}\ }\textbf {\bibinfo {volume} {9}},\ \bibinfo {pages}
  {663} (\bibinfo {year} {2007})}\BibitemShut {NoStop}%
\bibitem [{\citenamefont {Succi}(2008)}]{succi2008lattice}%
  \BibitemOpen
  \bibfield  {author} {\bibinfo {author} {\bibfnamefont {S.}~\bibnamefont
  {Succi}},\ }\href@noop {} {\bibfield  {journal} {\bibinfo  {journal} {The
  European Physical Journal B-Condensed Matter and Complex Systems}\ }\textbf
  {\bibinfo {volume} {64}},\ \bibinfo {pages} {471} (\bibinfo {year}
  {2008})}\BibitemShut {NoStop}%
\bibitem [{\citenamefont {Shan}\ and\ \citenamefont
  {He}(1998)}]{shan1998discretization}%
  \BibitemOpen
  \bibfield  {author} {\bibinfo {author} {\bibfnamefont {X.}~\bibnamefont
  {Shan}}\ and\ \bibinfo {author} {\bibfnamefont {X.}~\bibnamefont {He}},\
  }\href@noop {} {\bibfield  {journal} {\bibinfo  {journal} {Phys. Rev. Lett.}\
  }\textbf {\bibinfo {volume} {80}},\ \bibinfo {pages} {65} (\bibinfo {year}
  {1998})}\BibitemShut {NoStop}%
\bibitem [{\citenamefont {Bhatnagar}\ \emph {et~al.}(1954)\citenamefont
  {Bhatnagar}, \citenamefont {Gross},\ and\ \citenamefont
  {Krook}}]{PhysRev.94.511}%
  \BibitemOpen
  \bibfield  {author} {\bibinfo {author} {\bibfnamefont {P.~L.}\ \bibnamefont
  {Bhatnagar}}, \bibinfo {author} {\bibfnamefont {E.~P.}\ \bibnamefont
  {Gross}}, \ and\ \bibinfo {author} {\bibfnamefont {M.}~\bibnamefont
  {Krook}},\ }\href {\doibase 10.1103/PhysRev.94.511} {\bibfield  {journal}
  {\bibinfo  {journal} {Phys. Rev.}\ }\textbf {\bibinfo {volume} {94}},\
  \bibinfo {pages} {511} (\bibinfo {year} {1954})}\BibitemShut {NoStop}%
\bibitem [{\citenamefont {Mendoza}\ \emph
  {et~al.}(2010{\natexlab{a}})\citenamefont {Mendoza}, \citenamefont
  {Boghosian}, \citenamefont {Herrmann},\ and\ \citenamefont
  {Succi}}]{PhysRevLett.105.014502}%
  \BibitemOpen
  \bibfield  {author} {\bibinfo {author} {\bibfnamefont {M.}~\bibnamefont
  {Mendoza}}, \bibinfo {author} {\bibfnamefont {B.~M.}\ \bibnamefont
  {Boghosian}}, \bibinfo {author} {\bibfnamefont {H.~J.}\ \bibnamefont
  {Herrmann}}, \ and\ \bibinfo {author} {\bibfnamefont {S.}~\bibnamefont
  {Succi}},\ }\href {\doibase 10.1103/PhysRevLett.105.014502} {\bibfield
  {journal} {\bibinfo  {journal} {Phys. Rev. Lett.}\ }\textbf {\bibinfo
  {volume} {105}},\ \bibinfo {pages} {014502} (\bibinfo {year}
  {2010}{\natexlab{a}})}\BibitemShut {NoStop}%
\bibitem [{\citenamefont {Mendoza}\ \emph
  {et~al.}(2010{\natexlab{b}})\citenamefont {Mendoza}, \citenamefont
  {Boghosian}, \citenamefont {Herrmann},\ and\ \citenamefont {Succi}}]{rlbPRD}%
  \BibitemOpen
  \bibfield  {author} {\bibinfo {author} {\bibfnamefont {M.}~\bibnamefont
  {Mendoza}}, \bibinfo {author} {\bibfnamefont {B.~M.}\ \bibnamefont
  {Boghosian}}, \bibinfo {author} {\bibfnamefont {H.~J.}\ \bibnamefont
  {Herrmann}}, \ and\ \bibinfo {author} {\bibfnamefont {S.}~\bibnamefont
  {Succi}},\ }\href {\doibase 10.1103/PhysRevD.82.105008} {\bibfield  {journal}
  {\bibinfo  {journal} {Phys. Rev. D}\ }\textbf {\bibinfo {volume} {82}},\
  \bibinfo {pages} {105008} (\bibinfo {year} {2010}{\natexlab{b}})}\BibitemShut
  {NoStop}%
\bibitem [{\citenamefont {Bouras}\ \emph {et~al.}(2009)\citenamefont {Bouras},
  \citenamefont {Molnar}, \citenamefont {Niemi}, \citenamefont {Xu},
  \citenamefont {El}, \citenamefont {Fochler}, \citenamefont {Greiner},\ and\
  \citenamefont {Rischke}}]{bouras2009relativistic}%
  \BibitemOpen
  \bibfield  {author} {\bibinfo {author} {\bibfnamefont {I.}~\bibnamefont
  {Bouras}}, \bibinfo {author} {\bibfnamefont {E.}~\bibnamefont {Molnar}},
  \bibinfo {author} {\bibfnamefont {H.}~\bibnamefont {Niemi}}, \bibinfo
  {author} {\bibfnamefont {Z.}~\bibnamefont {Xu}}, \bibinfo {author}
  {\bibfnamefont {A.}~\bibnamefont {El}}, \bibinfo {author} {\bibfnamefont
  {O.}~\bibnamefont {Fochler}}, \bibinfo {author} {\bibfnamefont
  {C.}~\bibnamefont {Greiner}}, \ and\ \bibinfo {author} {\bibfnamefont
  {D.}~\bibnamefont {Rischke}},\ }\href@noop {} {\bibfield  {journal} {\bibinfo
   {journal} {Phys. Rev. Lett.}\ }\textbf {\bibinfo {volume} {103}},\ \bibinfo
  {pages} {32301} (\bibinfo {year} {2009})}\BibitemShut {NoStop}%
\bibitem [{\citenamefont {Li}\ \emph {et~al.}(2012)\citenamefont {Li},
  \citenamefont {Luo},\ and\ \citenamefont {Li}}]{mrtrlbPRD}%
  \BibitemOpen
  \bibfield  {author} {\bibinfo {author} {\bibfnamefont {Q.}~\bibnamefont
  {Li}}, \bibinfo {author} {\bibfnamefont {K.~H.}\ \bibnamefont {Luo}}, \ and\
  \bibinfo {author} {\bibfnamefont {X.~J.}\ \bibnamefont {Li}},\ }\href
  {\doibase 10.1103/PhysRevD.86.085044} {\bibfield  {journal} {\bibinfo
  {journal} {Phys. Rev. D}\ }\textbf {\bibinfo {volume} {86}},\ \bibinfo
  {pages} {085044} (\bibinfo {year} {2012})}\BibitemShut {NoStop}%
\bibitem [{\citenamefont {Romatschke}\ \emph {et~al.}(2011)\citenamefont
  {Romatschke}, \citenamefont {Mendoza},\ and\ \citenamefont
  {Succi}}]{PhysRevC.84.034903}%
  \BibitemOpen
  \bibfield  {author} {\bibinfo {author} {\bibfnamefont {P.}~\bibnamefont
  {Romatschke}}, \bibinfo {author} {\bibfnamefont {M.}~\bibnamefont {Mendoza}},
  \ and\ \bibinfo {author} {\bibfnamefont {S.}~\bibnamefont {Succi}},\ }\href
  {\doibase 10.1103/PhysRevC.84.034903} {\bibfield  {journal} {\bibinfo
  {journal} {Phys. Rev. C}\ }\textbf {\bibinfo {volume} {84}},\ \bibinfo
  {pages} {034903} (\bibinfo {year} {2011})}\BibitemShut {NoStop}%
\bibitem [{\citenamefont {Anderson}\ and\ \citenamefont
  {Witting}(1974)}]{Anderson1974466}%
  \BibitemOpen
  \bibfield  {author} {\bibinfo {author} {\bibfnamefont {J.}~\bibnamefont
  {Anderson}}\ and\ \bibinfo {author} {\bibfnamefont {H.}~\bibnamefont
  {Witting}},\ }\href {\doibase 10.1016/0031-8914(74)90355-3} {\bibfield
  {journal} {\bibinfo  {journal} {Physica}\ }\textbf {\bibinfo {volume} {74}},\
  \bibinfo {pages} {466 } (\bibinfo {year} {1974})}\BibitemShut {NoStop}%
\bibitem [{\citenamefont {Marle}\ and\ \citenamefont {Hebad}(1965)}]{marle}%
  \BibitemOpen
  \bibfield  {author} {\bibinfo {author} {\bibfnamefont {C.}~\bibnamefont
  {Marle}}\ and\ \bibinfo {author} {\bibfnamefont {C.}~\bibnamefont {Hebad}},\
  }\href@noop {} {\bibfield  {journal} {\bibinfo  {journal} {Seances Acad.
  Sci.}\ }\textbf {\bibinfo {volume} {260}},\ \bibinfo {pages} {6539} (\bibinfo
  {year} {1965})}\BibitemShut {NoStop}%
\bibitem [{\citenamefont {Stewart}(1971)}]{stewart}%
  \BibitemOpen
  \bibfield  {author} {\bibinfo {author} {\bibfnamefont {J.}~\bibnamefont
  {Stewart}},\ }\href@noop {} {\emph {\bibinfo {title} {Non-equilibrium
  Relativistic Kinetic Theory}}}\ (\bibinfo  {publisher} {Springer},\ \bibinfo
  {address} {Berlin},\ \bibinfo {year} {1971})\BibitemShut {NoStop}%
\bibitem [{\citenamefont {Cercignani}\ and\ \citenamefont
  {Kremer}(2002)}]{cercignani}%
  \BibitemOpen
  \bibfield  {author} {\bibinfo {author} {\bibfnamefont {C.}~\bibnamefont
  {Cercignani}}\ and\ \bibinfo {author} {\bibfnamefont {G.}~\bibnamefont
  {Kremer}},\ }\href@noop {} {\emph {\bibinfo {title} {The Relativistic
  Boltzmann Equation: Theory and Apllications}}}\ (\bibinfo  {publisher}
  {Birkhauser},\ \bibinfo {address} {Boston; Basel; Berlin},\ \bibinfo {year}
  {2002})\BibitemShut {NoStop}%
\bibitem [{\citenamefont {Hupp}\ \emph {et~al.}(2011)\citenamefont {Hupp},
  \citenamefont {Mendoza}, \citenamefont {Bouras}, \citenamefont {Succi},\ and\
  \citenamefont {Herrmann}}]{hupp2011relativistic}%
  \BibitemOpen
  \bibfield  {author} {\bibinfo {author} {\bibfnamefont {D.}~\bibnamefont
  {Hupp}}, \bibinfo {author} {\bibfnamefont {M.}~\bibnamefont {Mendoza}},
  \bibinfo {author} {\bibfnamefont {I.}~\bibnamefont {Bouras}}, \bibinfo
  {author} {\bibfnamefont {S.}~\bibnamefont {Succi}}, \ and\ \bibinfo {author}
  {\bibfnamefont {H.}~\bibnamefont {Herrmann}},\ }\href@noop {} {\bibfield
  {journal} {\bibinfo  {journal} {Physical Review D}\ }\textbf {\bibinfo
  {volume} {84}},\ \bibinfo {pages} {125015} (\bibinfo {year}
  {2011})}\BibitemShut {NoStop}%
\bibitem [{\citenamefont {Pan}\ \emph {et~al.}(2007)\citenamefont {Pan},
  \citenamefont {Xu}, \citenamefont {Zhang},\ and\ \citenamefont
  {Jiang}}]{pan2007lattice}%
  \BibitemOpen
  \bibfield  {author} {\bibinfo {author} {\bibfnamefont {X.}~\bibnamefont
  {Pan}}, \bibinfo {author} {\bibfnamefont {A.}~\bibnamefont {Xu}}, \bibinfo
  {author} {\bibfnamefont {G.}~\bibnamefont {Zhang}}, \ and\ \bibinfo {author}
  {\bibfnamefont {S.}~\bibnamefont {Jiang}},\ }\href@noop {} {\bibfield
  {journal} {\bibinfo  {journal} {International Journal of Modern Physics C}\
  }\textbf {\bibinfo {volume} {18}},\ \bibinfo {pages} {1747} (\bibinfo {year}
  {2007})}\BibitemShut {NoStop}%
\bibitem [{\citenamefont {Dellar}(2001)}]{dellar2001bulk}%
  \BibitemOpen
  \bibfield  {author} {\bibinfo {author} {\bibfnamefont {P.}~\bibnamefont
  {Dellar}},\ }\href@noop {} {\bibfield  {journal} {\bibinfo  {journal}
  {Physical Review E}\ }\textbf {\bibinfo {volume} {64}},\ \bibinfo {pages}
  {031203} (\bibinfo {year} {2001})}\BibitemShut {NoStop}%
\bibitem [{\citenamefont {Romatschke}(2010)}]{romatschke2010new}%
  \BibitemOpen
  \bibfield  {author} {\bibinfo {author} {\bibfnamefont {P.}~\bibnamefont
  {Romatschke}},\ }\href@noop {} {\bibfield  {journal} {\bibinfo  {journal}
  {International Journal of Modern Physics E}\ }\textbf {\bibinfo {volume}
  {19}},\ \bibinfo {pages} {1} (\bibinfo {year} {2010})}\BibitemShut {NoStop}%
\bibitem [{\citenamefont {Mendoza}\ \emph {et~al.}(2013)\citenamefont
  {Mendoza}, \citenamefont {Karlin}, \citenamefont {Succi},\ and\ \citenamefont
  {Herrmann}}]{mendoza2013ultrarelativistic}%
  \BibitemOpen
  \bibfield  {author} {\bibinfo {author} {\bibfnamefont {M.}~\bibnamefont
  {Mendoza}}, \bibinfo {author} {\bibfnamefont {I.}~\bibnamefont {Karlin}},
  \bibinfo {author} {\bibfnamefont {S.}~\bibnamefont {Succi}}, \ and\ \bibinfo
  {author} {\bibfnamefont {H.}~\bibnamefont {Herrmann}},\ }\href@noop {}
  {\bibfield  {journal} {\bibinfo  {journal} {arXiv:1301.3420 to appear in
  JSTAT}\ } (\bibinfo {year} {2013})}\BibitemShut {NoStop}%
\bibitem [{\citenamefont {Mei}\ and\ \citenamefont
  {Shyy}(1998)}]{mei1998finite}%
  \BibitemOpen
  \bibfield  {author} {\bibinfo {author} {\bibfnamefont {R.}~\bibnamefont
  {Mei}}\ and\ \bibinfo {author} {\bibfnamefont {W.}~\bibnamefont {Shyy}},\
  }\href@noop {} {\bibfield  {journal} {\bibinfo  {journal} {Journal of
  Computational Physics}\ }\textbf {\bibinfo {volume} {143}},\ \bibinfo {pages}
  {426} (\bibinfo {year} {1998})}\BibitemShut {NoStop}%
\bibitem [{\citenamefont {Rischke}\ \emph {et~al.}(1995)\citenamefont
  {Rischke}, \citenamefont {Bernard},\ and\ \citenamefont
  {Maruhn}}]{Rischke1995346}%
  \BibitemOpen
  \bibfield  {author} {\bibinfo {author} {\bibfnamefont {D.~H.}\ \bibnamefont
  {Rischke}}, \bibinfo {author} {\bibfnamefont {S.}~\bibnamefont {Bernard}}, \
  and\ \bibinfo {author} {\bibfnamefont {J.~A.}\ \bibnamefont {Maruhn}},\
  }\href {\doibase 10.1016/0375-9474(95)00355-1} {\bibfield  {journal}
  {\bibinfo  {journal} {Nuclear Physics A}\ }\textbf {\bibinfo {volume}
  {595}},\ \bibinfo {pages} {346 } (\bibinfo {year} {1995})}\BibitemShut
  {NoStop}%
\bibitem [{\citenamefont {Thompson}(1986)}]{thompson1986special}%
  \BibitemOpen
  \bibfield  {author} {\bibinfo {author} {\bibfnamefont {K.}~\bibnamefont
  {Thompson}},\ }\href@noop {} {\bibfield  {journal} {\bibinfo  {journal}
  {Journal of Fluid Mechanics}\ }\textbf {\bibinfo {volume} {171}},\ \bibinfo
  {pages} {365} (\bibinfo {year} {1986})}\BibitemShut {NoStop}%
\bibitem [{\citenamefont {Nadiga}(1995)}]{nadiga1995euler}%
  \BibitemOpen
  \bibfield  {author} {\bibinfo {author} {\bibfnamefont {B.}~\bibnamefont
  {Nadiga}},\ }\href@noop {} {\bibfield  {journal} {\bibinfo  {journal}
  {Journal of statistical physics}\ }\textbf {\bibinfo {volume} {81}},\
  \bibinfo {pages} {129} (\bibinfo {year} {1995})}\BibitemShut {NoStop}%
\bibitem [{\citenamefont {McKee}\ and\ \citenamefont
  {Draine}(1991)}]{scienceSuper}%
  \BibitemOpen
  \bibfield  {author} {\bibinfo {author} {\bibfnamefont {C.}~\bibnamefont
  {McKee}}\ and\ \bibinfo {author} {\bibfnamefont {B.}~\bibnamefont {Draine}},\
  }\href@noop {} {\bibfield  {journal} {\bibinfo  {journal} {Science}\ }\textbf
  {\bibinfo {volume} {252}},\ \bibinfo {pages} {397} (\bibinfo {year}
  {1991})}\BibitemShut {NoStop}%
\bibitem [{\citenamefont {$\mbox{A. M. Soderberg {\it et
  al.}}$}(2010)}]{natureSuper}%
  \BibitemOpen
  \bibfield  {author} {\bibinfo {author} {\bibnamefont {$\mbox{A. M. Soderberg
  {\it et al.}}$}},\ }\href@noop {} {\bibfield  {journal} {\bibinfo  {journal}
  {Nature Letters}\ }\textbf {\bibinfo {volume} {463}},\ \bibinfo {pages} {513}
  (\bibinfo {year} {2010})}\BibitemShut {NoStop}%
\end{thebibliography}%

\end{document}